\documentclass[prd,aps,twocolumn,floatfix,superscriptaddress]{revtex4}

\usepackage{amssymb,graphicx}
\usepackage{epsfig}

\def\rmd{{\rm d}}

\font\bfgreek=cmmib10

\newcommand{\had}{{\sc had}}

\def\bbf{{\hbox{\bfgreek\char'146}}}

\def\bbq{{\hbox{\bfgreek\char'161}}}

\def\bbs{{\hbox{\bfgreek\char'163}}}

\def\bbu{{\hbox{\bfgreek\char'165}}}

\def\bbw{{\hbox{\bfgreek\char'167}}}

\begin{document}

\title{Simulating binary neutron stars: dynamics and gravitational waves}

\author{Matthew Anderson}
\affiliation{Department of Physics and Astronomy, Louisiana State
University, Baton Rouge, LA 70803-4001}
\author{Eric W. Hirschmann} 
\affiliation{Department of Physics and Astronomy, Brigham Young
University, Provo, UT 84602}
\author{Luis Lehner}
\affiliation{Department of Physics and Astronomy, Louisiana State
University, Baton Rouge, LA 70803-4001}
\author{Steven L. Liebling}
\affiliation{Department of Physics, Long Island University -- C.W. Post Campus,
Brookville, NY 11548}
\author{Patrick M. Motl} 
\affiliation{Department of Physics and Astronomy, Louisiana State
University, Baton Rouge, LA 70803-4001}
\author{David Neilsen}
\affiliation{Department of Physics and Astronomy, Brigham Young
University, Provo, UT 84602}
\author{Carlos Palenzuela} 
\affiliation{Department of Physics and Astronomy, Louisiana State
University, Baton Rouge, LA 70803-4001}
\author{Joel E. Tohline}
\affiliation{Department of Physics and Astronomy, Louisiana State
University, Baton Rouge, LA 70803-4001}

\date{\today}

%
%
\begin{abstract}
We model two mergers of orbiting binary neutron stars, the first forming 
a black hole and the second a differentially rotating neutron star.  We extract 
gravitational waveforms in the wave zone.  Comparisons to a post-Newtonian
analysis allow us to compute the orbital kinematics, including trajectories
and orbital eccentricities.  We verify our code by evolving single stars 
and extracting radial perturbative modes, which compare very well to
results from perturbation theory.  The Einstein equations are solved in
a first order reduction of the generalized harmonic formulation, and 
the fluid equations are solved using a modified convex essentially 
non-oscillatory method. All calculations are done in three spatial dimensions without symmetry assumptions.
We use the \had\ computational infrastructure for distributed
adaptive mesh refinement.
\end{abstract}

\maketitle

%
%
\section{Introduction}

It is widely expected that gravitational waves of sufficiently strong 
amplitude will 
be detected by a new generation of gravitational wave interferometers. 
Binary systems composed of
compact objects, such as black holes and/or neutron stars, are among
the strongest expected sources of these waves.
Advanced gravitational wave detectors should be sensitive enough to
detect the merging phase of such binaries.  A detailed analysis 
of the expected waveforms from these events 
will provide valuable information not only in the analysis of the
received signals, but also in the design and tuning of future advanced 
gravitational wave detectors~\cite{advligo,Mandel:2007hi}.

In relation to these efforts, and beyond the intrinsic importance of 
the two-body problem in general relativity (GR), it is significant that recent studies of 
the binary black hole problem have made substantial progress in providing waveforms for
these mergers (see for instance~\cite{Pret05,CLMZ06,BCCKM06,DHPS06,GSBHH07,SPRTW07}).
Furthermore, these numerical results for vacuum spacetimes show a remarkable
agreement with those obtained with approximation techniques \cite{BCP07,PBBC07}.  
This provides considerable support for the use of
waveforms obtained via approximation techniques, suitably enhanced by further information
from numerical simulations, since these can be more
easily encoded in a template bank~\cite{Baumgarte:2006en}. This requires knowing the waveforms during the
pre-merger, merger and post-merger stages and matching them appropriately to obtain
the continuous wavetrain through the most violent and strongly radiative stage of
the dynamics.  

For non-vacuum spacetimes, differences in the waveforms 
may arise from the state of matter describing the compact stars, the 
influence of magnetic fields and related phenomena.
To fully understand these systems and their waveforms,
detailed simulations will be required to map out the possible phenomenology. 

For the particular case of binary neutron
stars in full GR, several efforts studying the system in three dimensional
settings have been presented in recent
years~\cite{Shibata:2002jb,Duez:2002bn,Shibata:2003ga,
Miller:2003vc,Marronetti:2003hx,Jin:2006gm,Shibata:2005ss}.  
However,
the complexity and computational cost of these simulations has 
permitted investigators to consider only a portion of the interesting parameter
space and several of them have been restricted by symmetry
considerations.  Nevertheless, a number of interesting problems are 
beginning to be addressed, including the 
influence of stiff versus soft equations of state~\cite{Shibata:2005ss},
a possible way to determine the innermost stable circular
orbit~\cite{Marronetti:2003hx}, the dynamics of unequal mass
binaries~\cite{Shibata:2005ss} and even the possible existence of
critical phenomena in the merging system~\cite{Jin:2006gm}.

Further exploration of these systems will require relaxing symmetry 
considerations, such as axisymmetry or equatorial symmetry, and expanding 
the space of initial configurations that can be successfully evolved.  
Moreover, the inclusion of additional physics such as
magnetic fields will be important as these effects may play a major role in
the resulting dynamics.  For instance, the magnetorotational instability, 
which redistributes angular momentum in the system, can have a strong 
influence on the multipole structure of the central source and
hence on the gravitational wave output of the system.

To date, work on black hole-neutron star binaries has been limited to a few
cases~\cite{Shibata:2006,Shibata:2006a,Rezzolla:2006}.  As a result, our
understanding of this type of system is still in its infancy.
Needless to say, we have even more to understand about 
both types of compact binaries when their environments, which may include 
magnetic fields and radiation transport, are included.  
Indeed, both magnetic fields and radiation transport are expected to be 
key ingredients in modeling short, hard gamma-ray burst phenomena with compact 
binaries.  
Understanding such spectacular events requires the addition of
  these ingredients to the computational infrastructure. The
  resulting numerical simulations should allow for new astrophysical
  insights.

The present work is intended as the first in a series of studies 
that examine the evolution of compact binary systems in full three 
dimensional general relativity. 
To this end, we have developed a general computational infrastructure
with solvers for the Einstein and relativistic MHD equations that 
incorporates several novel
features, which we discuss in the following sections.  In Section II 
we describe our formulation of the equations for these systems.
This includes expressing 
the Einstein equations in terms of a desirable symmetric hyperbolic property~\cite{Palenzuela:2006wp,PLLinprep}
and coupling them with the equations of relativistic magnetohydrodynamics (MHD)~\cite{Neilsen2005,Anderson:2006ay}. Section III presents our
numerical implementation, such as integration techniques,
distributed adaptive mesh refinement (AMR), and 
a tapered grid algorithm that ensures 
stability and considerably reduces spurious reflections off artificial
internal 
boundaries~\cite{Lehner:2005vc}.  
These ingredients let
us simulate binary evolutions in which the stars begin with wider
separations than has been done in earlier studies.  We can
extract gravitational radiation in the wave zone and
place outer boundaries an order of magnitude beyond what has been done
previously.
As a result, contamination by boundary effects
is negligible.
Section IV presents a fairly stringent code test by considering the 
dynamics of a single
Tolman-Oppenheimer-Volkoff (TOV) solution and extracting the
radial oscillation modes of the star.  
Section V describes our main application, namely a study of a binary 
neutron star system without any assumed symmetries. We follow
the dynamics of the system from an early non-quasicircular stage
to the merger and subsequent formation of a neutron star or a black
hole.  We present gravitational wave signals as measured by observers 
placed in the wave-zone and calculated via Weyl scalars.  
Section VI concludes and offers some considerations for future 
work.

%
%

\section{Formulation and equations of motion}

The binary neutron star systems considered here are governed by both
the Einstein equations for the geometry, and the relativistic fluid
equations for the matter.  We write both systems as first order
hyperbolic equations.

In this section we present a brief summary of our formulation and equations
for both the geometry and the fluid.  More details on our 
approach to the Einstein equations~\cite{Palenzuela:2006wp} and the
relativistic fluid equations~\cite{Neilsen2005,Anderson:2006ay} can be
found elsewhere.
By way of notation, we use letters from the beginning of the 
alphabet
($a$, $b$, $c$) for spacetime indices, while letters from the middle
of the alphabet ($i$, $j$, $k$) range over spatial components.
We adopt geometric units where $c=G=1$.

\subsection{Einstein equations}

We write the Einstein equations in a first order reduction of the
generalized harmonic (GH) formalism.  Our approach is closely related to 
the one in~\cite{Lindblom:2005qh}, and it was used 
previously in binary boson star evolutions~\cite{Palenzuela:2006wp},
where additional information can be found.

We define spacelike hypersurfaces at
$x^0\equiv t ={\rm const.}$, and define the 3-metric $h_{ij}$ on the
hypersurfaces.  A vector normal to the hypersurfaces is given by
$n_a = - \nabla_a t / ||\nabla_a t ||$, and coordinates defined on 
neighboring hypersurfaces can be related through the lapse, $\alpha$, 
and shift, $\beta^i$.  With these definitions, the spacetime metric 
$g_{ab}$ can then be written as
\begin{eqnarray}
\rmd s^2 &=& g_{ab}\, \rmd x^a \rmd x^b\\
         &=&-\alpha^2 \, \rmd t^2 
            + h_{ij}\left(\rmd x^i + \beta^i\, \rmd t\right)
                    \left(\rmd x^j + \beta^j\, \rmd t\right).
\end{eqnarray}
Indices on spacetime quantities are raised and lowered with the 4-metric,
$g_{ab}$, and its inverse, while the 3-metric $h_{ij}$ and its inverse
are used to raise and lower indices on spatial quantities.

In the generalized harmonic formulation, the evolved variables are
\begin{equation}
 g_{ab} \, , \quad Q_{ab} \equiv - n^c \, \partial_c g_{ab} \, ,
\quad  D_{iab} \equiv \partial_i g_{ab} \, ,
\end{equation}
namely the spacetime metric and its temporal and spatial derivatives, 
respectively.  Coordinates are specified via the generalized harmonic 
condition
\begin{equation}
\square x^a = H^a(t,x^i),
\end{equation}
where the arbitrary source 
functions $H^a(t,x^i)$ determine the coordinate freedom.  
Although our code allows for a general coordinate choice, we
choose harmonic coordinates for the work presented here and set
$H^a(t,x^i)=0$.

The evolution equations in our GH formalism are
\begin{eqnarray}
  \label{EE_geq}
  \partial_t g_{ab} &=& \beta^k~D_{kab} - \alpha~Q_{ab}, \\
  \partial_t Q_{ab} &=& \beta^k~\partial_k Q_{ab}
  - \alpha h^{ij} \partial_i D_{jab} \nonumber \\ 
  &-& \alpha~ \partial_a H_b - \alpha~ \partial_b H_a +
  2~\alpha~ \Gamma_{cab}~ H^c \nonumber \\
  &+& 2\, \alpha\, g^{cd}~(h^{ij} D_{ica} D_{jdb} - Q_{ca} Q_{db}
                   - g^{ef} \Gamma_{ace} \Gamma_{bdf}) \nonumber \\
  &-& \frac{\alpha}{2} n^c n^d Q_{cd} Q_{ab}
  - \alpha~h^{ij} D_{iab} Q_{jc} n^c \nonumber \\ 
  &-& 8 \pi \, \alpha(2T_{ab} - g_{ab} T) \nonumber \\
  &-& 2 \sigma_0 \, \alpha \, [n_a Z_b + n_b Z_a - g_{a b} n^c Z_c ]  \nonumber  \\
  &+& \sigma_1 \, \beta^i ( D_{iab} -  \partial_i g_{ab} ),  \\
   \label{EE_Deq}
   \partial_t D_{iab} &=& \beta^k \partial_k D_{iab}
  - \alpha~\partial_i Q_{ab} \nonumber \\ 
   &+& \frac{\alpha}{2} n^c n^d D_{icd} Q_{ab}
  + \alpha~h^{jk} n^c D_{ijc} D_{kab} \nonumber \\ 
   &-& \sigma_1 \, \alpha \, ( D_{iab} - \partial_i g_{ab} ) .
\end{eqnarray}
Here $T_{ab}$ is the stress-energy tensor and $T$ is its trace, $T=T^a{}_a$.
$Z^a$ is a vector related to the constraints defined below in 
Eq.~(\ref{eq:define_Z}). These variables are not evolved,
rather they measure the constraint violation and are included
in the evolution equations for constraint
 damping purposes~\cite{gundlach}.
We also define $\Gamma_{abc} = g_{ad}\Gamma^{d}{}_{bc}$, where
${\Gamma^a}{}_{bc}$ are the Christoffel symbols obtained from $g_{ab}$, 
given by
\begin{equation}
  \Gamma^a{}_{bc} = \frac{1}{2} \, {g^{ad}} (D_{bdc} + D_{cdb} - D_{dbc}) ~.
\end{equation}
Note, $D_{iab}$ are evolved variables in our system, and the
quantities $D_{0ab}$ are computed from $Q_{ab}$ and $D_{iab}$ via 
\begin{equation}
  D_{0ab} = -\alpha Q_{ab} + \beta^k D_{kab} \, .
\end{equation}
While the Arnowitt-Deser-Misner (ADM) extrinsic 
curvature is not part of the GH system, 
the fluid equations below are written in terms of $K_{ij}$, which can be
calculated as
\begin{equation}
  K_{ij} = \frac{1}{2} Q_{ij}+\frac{1}{\alpha}(D_{(ij)0}-\beta^k D_{(ij)k}).
\end{equation}

This GH formulation includes a number of constraints that must be satisfied 
for consistency, including the Hamiltonian and momentum constraints as well
as additional constraints that arise in the first order reduction.
In particular, if we define the four-vector
\begin{equation}
\label{harmonicZ}
    2 Z^a \equiv - \Gamma^{a}{}_{bc} \, g^{bc} - H^a(t,x^i) \, ,   
\label{eq:define_Z}
\end{equation}
it can be shown that the energy and momentum constraints are satisfied if 
$Z^a=0=\partial_t Z^a$.  
The free parameters $\sigma_0$ and $\sigma_1$ are chosen to 
control the damping of the four vector $Z_a$ (the
energy and momentum constraints) and the first order constraints,
respectively~\cite{Lindblom:2005qh,Palenzuela:2006wp}.  
We monitor the
$Z^a$ during the evolution as an indication of the magnitude of the numerical
error in the solution.

%
%
\subsection{Perfect fluid equations}

We now briefly introduce the perfect fluid equations.  
Additional information can be found in our previous 
work~\cite{Neilsen2005,Anderson:2006ay}
as well as in general review articles~\cite{Marti:1999wi,Font:2000pp}.

The stress-energy tensor for the perfect fluid is
\begin{equation}
T_{ab} = h_e u_a u_b + P g_{ab},
\end{equation}
where $u^a$ is the four velocity of the fluid, 
$h_e$ is the enthalpy, and $P$ is the isotropic pressure.  The enthalpy 
can be written
\begin{equation}
h_e = \rho_o + \rho_o\epsilon + P,
\end{equation}
where $\rho_o$ the rest energy density, and $\epsilon$ is
the specific internal energy density.
We introduce the quantities
\begin{equation}
W\equiv -n^a u_a, \qquad v^i \equiv \frac{1}{W}\,h^i{}_j u^j,
\end{equation}
where $W$ is the Lorentz factor between the fluid frame and the fiducial ADM
observers and $v^i$ is the spatial coordinate velocity of the fluid.
The set of fluid variables introduced here are known as the {\em primitive}
variables, ${\bbw} = (\rho_o , v^i , P)^{\rm T}$.

High resolution shock capturing schemes (HRSC) are robust numerical methods for
compressible fluid dynamics.  These methods, based on Godunov's seminal
work~\cite{Godunov}, are fundamentally based on writing the fluid equations as
integral conservation laws.  To this end, we introduce {\em conservative}
variables ${\bbq} = (D, S_i, \tau)^{\rm T}$, where
\begin{eqnarray}
D &=& W \rho_o,\\
S_i &=& h_e W^2 v_i,\\
\tau &=& h_e W^2 - P - D.
\end{eqnarray}
In an asymptotically flat spacetime these quantities are conserved, 
and are related to the
baryon number, momentum, and, in the non-relativistic limit, the kinetic
energy, respectively.  Anticipating the form of the evolution equations,
we also introduce the densitized conserved variables
\begin{equation}
\tilde D = \sqrt{h}\, D, \quad 
\tilde S_i = \sqrt{h}\, S_i, \quad
\tilde \tau = \sqrt{h}\, \tau,
\end{equation}
where $h=\det(h_{ij})$. 
The fluid equations can now be written in balance law form
\begin{equation}
\partial_t\tilde\bbq  + \partial_k\bbf\,^k(\tilde\bbq) = \bbs(\tilde\bbq),
\label{eq:balance}
\end{equation}
where $\bbf\,^k$ are flux functions, and $\bbs$
are source terms.  The fluid equations in this form are specifically
\begin{eqnarray}
&&\partial_t \tilde D + \partial_i \left[ \alpha\,\tilde D
\left( v^i - {\beta^i \over \alpha} \right) \right] = 0,\label{eq:ev_D} \\
&& \partial_t \tilde S_j + \partial_i \left[ \alpha \left(
\tilde S_j \left( v^i - {\beta^i \over \alpha} \right) + \sqrt{h}\,P \, h^i{}_j
   \right)\right]\nonumber\\
&&\qquad  = \alpha \,  {^{3}{\Gamma}}^i{}_{jk} \, \left( \tilde S_i v^k 
         + \sqrt{h}\,P h_i{}^k \right)  
   + \tilde S_a\partial_j\beta^a\Bigr.\nonumber\\
&&\qquad\qquad\qquad
  - \partial_j \alpha \, (\tilde\tau + \tilde D), \\
 &&\partial_t \tilde\tau
        + \partial_i \left[ \alpha\left(\tilde S^i - \frac{\beta^i}{\alpha} \, 
                     \tilde\tau - v^i \tilde D \right) \right] \nonumber\\
&& \qquad= \alpha \, 
\left[ K_{ij} \tilde S^i v^j + \sqrt{h}\, K P - \frac{1}{\alpha} 
\, \tilde S^a \partial_a \alpha \right].
\end{eqnarray}
Here ${^{3}{\Gamma}}^i{}_{jk}$ is the Christoffel symbol associated with the
3-metric $h_{ij}$, and $K$ is the trace of the extrinsic curvature, 
$K = K^i{}_i$.

Finally, we close the system of fluid equations with an equation of state (EOS).
We choose the ideal gas EOS
\begin{equation}
P = (\Gamma-1)\, \rho_o\epsilon,
\end{equation}
where $\Gamma$ is the constant adiabatic exponent.  Nuclear matter in 
neutron stars
is relatively stiff, and we set $\Gamma=2$ in this work.
When the fluid flow is adiabatic, this EOS reduces to the well known
polytropic EOS 
\begin{equation}
P=\kappa\rho_o{}^\Gamma,
\label{eq:polyEOS}
\end{equation}
where $\kappa$ is a dimensional constant.   We use the polytropic EOS
only for setting initial data.

%
%

\section{Numerical approach}

Our numerical approach to solving the combined equations of general 
relativistic hydrodynamics (GRHD) is built upon two extensively tested
codes:  These were written to solve the (1) Einstein 
equations~\cite{Palenzuela:2006wp,PLLinprep}, and the (2) relativistic
magnetohydrodynamics (MHD) equations~\cite{Neilsen2005,Anderson:2006ay}.
It should be mentioned that while we do solve the full GRMHD equations, in 
our current work the magnetic field is set to zero.  
Results with non-trivial magnetic fields will
be presented elsewhere~\cite{everybodyPRL}.  

While both sets of evolution
equations are hyperbolic, the solutions from each set of equations are quite
different.  The Einstein equations are linearly degenerate, and 
therefore we expect smooth solutions to evolve from smooth initial data.
The fluid equations, on the other hand, are genuinely nonlinear, and
discontinuous weak solutions (shocks) generically evolve from smooth initial 
data~\cite{Reula:1998ty}.  We choose numerical methods adapted to the
features of each set of equations.
The fluid equations are evolved with a modified convex essentially 
non-oscillatory (CENO) method, while the Einstein equations are evolved
using fourth-order accurate difference operators that satisfy 
summation by parts (SBP). 
These very different methods are easily combined by discretizing the
equations in time using the method of lines.  

We base our code on the \had\ computational infrastructure for
distributed AMR.  The Einstein and fluid solvers are written in separate
modules, which can be used individually or combined.
The following sections
review our methods.

\subsection{Adaptive mesh refinement using \had}

The neutron star problem has several important physical scales, and
each must be adequately resolved to capture the relevant dynamics.
These scales include (1) the individual stars, preferably incorporating 
some of their internal dynamics, (2) the orbital
length scale, (3) the gravitational wave zone, and (4) the
location of outer boundaries.  In this work, the initial orbital scale 
is on the order of several stellar radii, the gravitational waves are
extracted at 30, 40 and 50 stellar radii, and the outer boundaries
of the computational domain are placed about 100 stellar radii from
the orbital pair to reduce boundary contamination of the orbital
dynamics and gravitational wave signals.  The computational demands
required to resolve these different physical scales are best met
using adaptive mesh refinement. 

\begin{figure}
\begin{center}
\epsfig{file=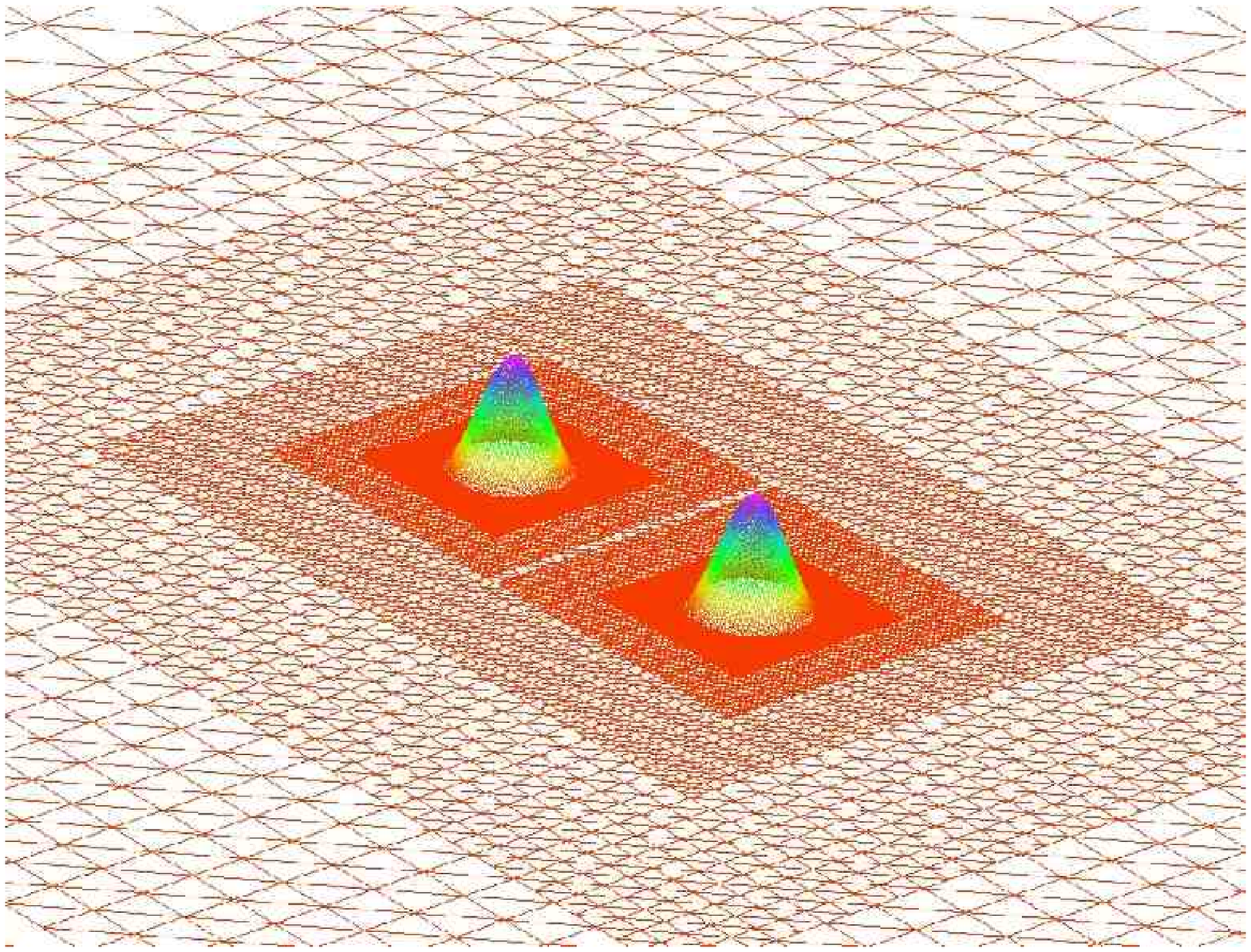,height=4.0cm}
\epsfig{file=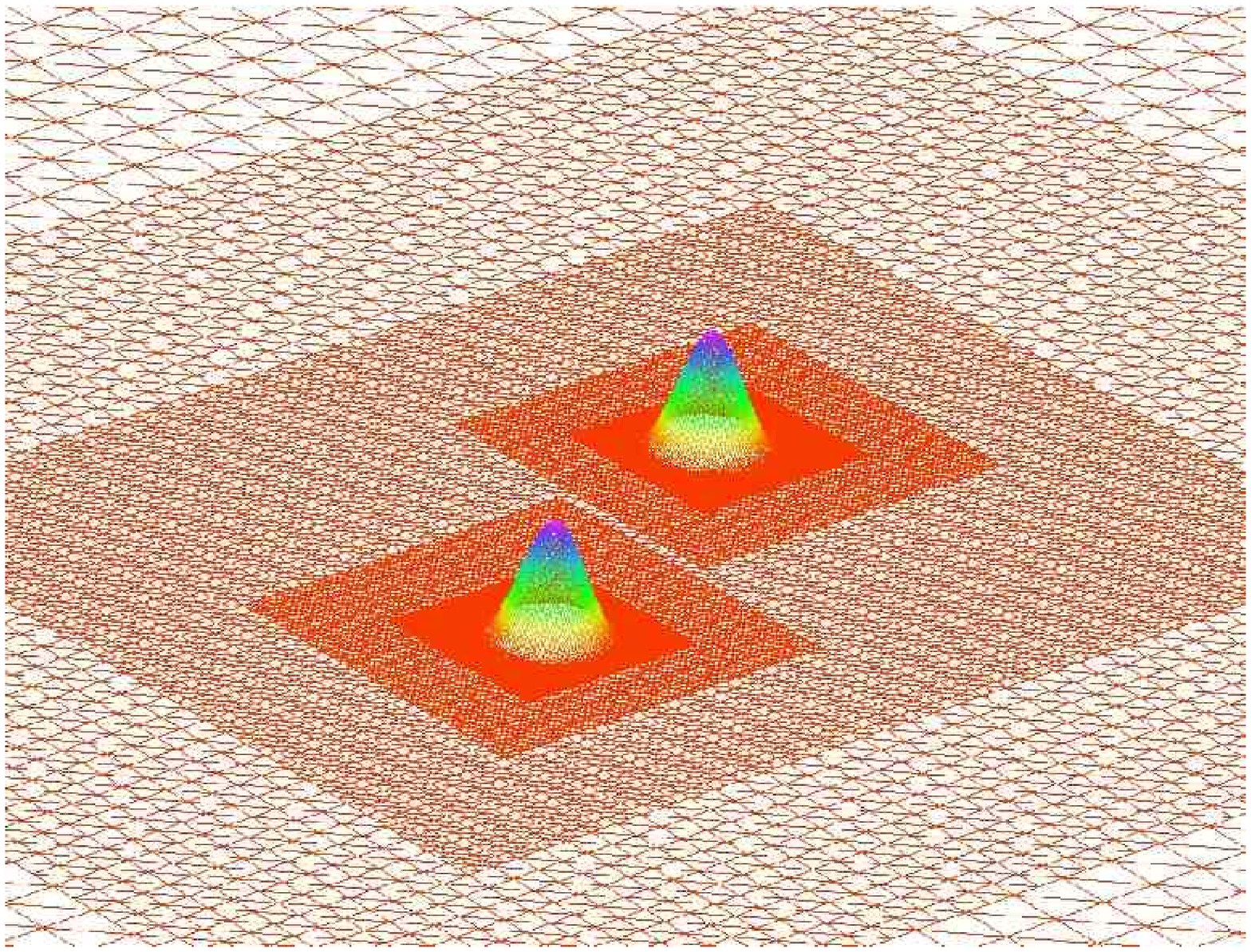,height=4.0cm}
\epsfig{file=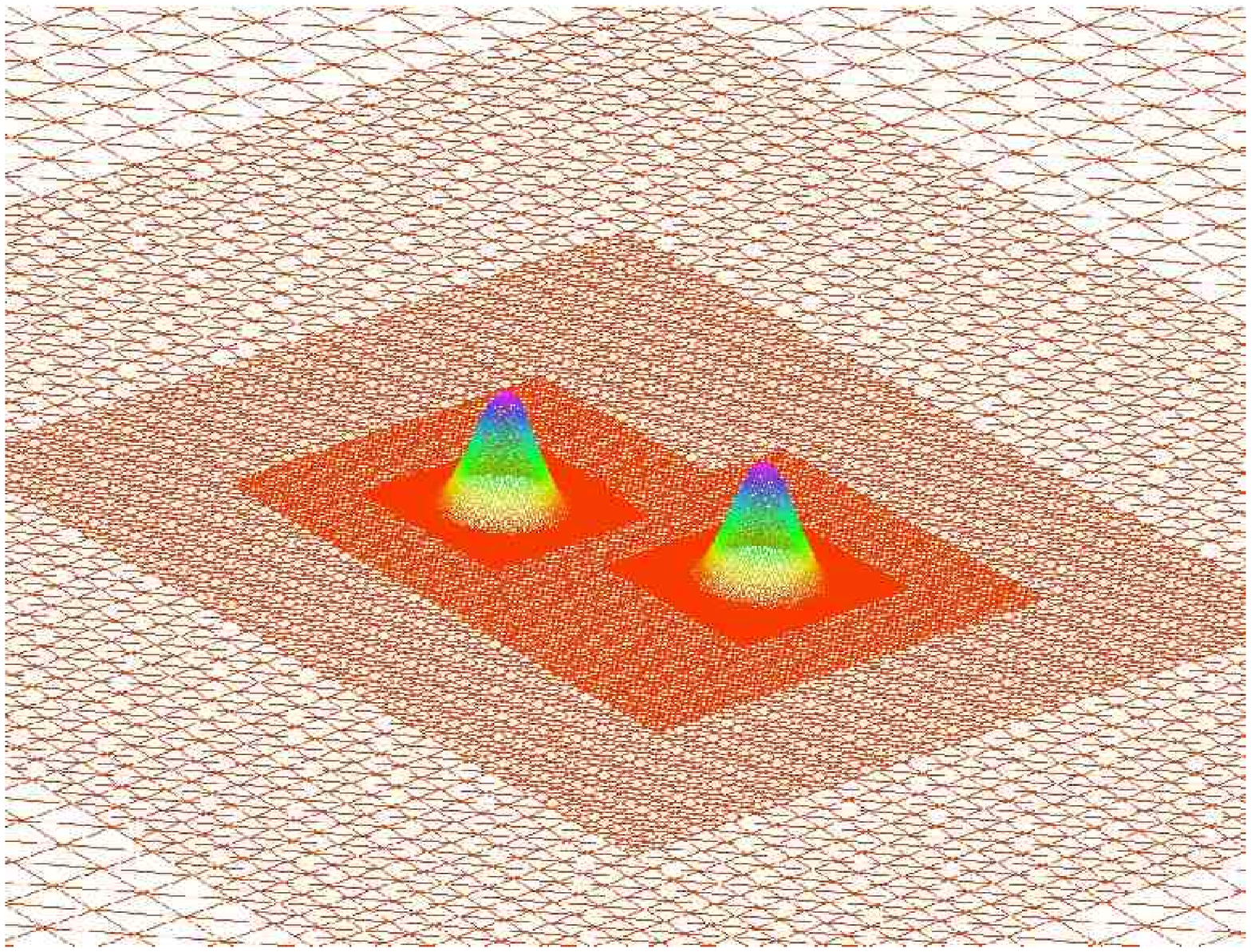,height=4.0cm}
\caption{The AMR mesh structure at times 0, 84, and 500 for the
    pre-merge stage of the simulation with 
      resolution of 32 points across each star.
    The simulation had seven levels of refinement, five of which are visible
     here.  Simulations were performed on 128 processors.} \label{fig:amr_mesh}
\end{center}
\end{figure}

We use the publicly available computational infrastructure 
\had\ to provide parallel distributed
AMR for our codes~\cite{had_webpage,Liebling}.  
\had\ can solve both hyperbolic and elliptic equations, and, unlike
several other publicly available AMR 
toolkits~\cite{SAMRAI,Chombo,Paramesh,AMROC,Boxlib}, it 
accommodates both  vertex and cell centered algorithms. \had\
has a modular design, allowing one to solve 
different sets of equations with the same computational infrastructure.
Furthermore, solvers for different equations can be coupled together,
as we have done here with separate solvers for the GR and MHD equations.
{\sc had} provides Berger-Oliger~\cite{Berger} style AMR with subcycling
in both space and time.  The {\sc had} clustering algorithm is
Berger-Rigoutsos~\cite{Rigoutsos}, and the load balancing algorithm is
the least loaded scheme~\cite{Rendleman}.  Refinement in \had\ can be
triggered by user-specified criteria, e.g., refining on solution
features such as gradients or extrema, or refining on truncation error
estimation using a shadow hierarchy.
The runs
presented here use the shadow hierarchy for refinement, and
all dynamic fields are used to estimate the truncation error. 
Some additional fixed refinement regions are used for gravitational wave 
extraction in the wave zone.
As an example, Figure~\ref{fig:amr_mesh} illustrates the resulting
mesh structure at a pre-merge stage in our simulations.

{\sc had} supports arbitrary orders of accuracy~\cite{Lehner:2005vc},
and the overall accuracy for the implementation employed here is third order for smooth
solutions.  {\sc had} implements the tapered-grid boundary method
for internal boundaries~\cite{Lehner:2005vc}.  This method is
advantageous for two reasons.  It guarantees stability of the AMR
algorithm if the unigrid counterpart is stable as well as significantly
reducing spurious reflections at interface boundaries.

Finally, when a fine grid is created during an evolution,
the geometric variables are interpolated onto the fine grid using
Lagrangian interpolation.  The fluid variables are interpolated using 
weighted essentially non-oscillatory (WENO)
interpolation~\cite{SebastianShu}.  
This interpolation scheme is designed for discontinuous
functions, and reduces to Lagrangian interpolation for smooth functions.

\subsection{Method of lines}

The numerical methods for the Einstein equations (SBP) and the
fluid equations (CENO) both specify the discretization of the spatial
difference operators, giving the semi-discrete equations
\begin{equation}
\frac{d \bbu}{dt} = L(\bbu).
\end{equation}
Here $\bbu$ represents the set of all variables evolved in both 
the Einstein and fluid equations, and $L$ represents a discrete spatial 
difference operator.  These ordinary differential equations are now
discretized in time using the method of lines.  
We choose a third order Runge-Kutta scheme that preserves the 
TVD (Total Variation
Diminishing) condition~\cite{ShuOsherI} to integrate the semi-discrete 
equations
\begin{eqnarray}
\bbu\,^{(1)} &=& \bbu\,^{n} + \triangle t L(\bbu\,^n),\nonumber\\
\bbu\,^{(2)} &=& \frac{3}{4}\bbu\,^{n} + \frac{1}{4}\bbu\,^{(1)} 
           + \frac{1}{4}\triangle t L(\bbu\,^{(1)}),\\
\bbu\,^{n+1} &=& \frac{1}{3}\bbu\,^{n} + \frac{2}{3}\bbu\,^{(2)} 
           + \frac{2}{3}\triangle t L(\bbu\,^{(2)}).\nonumber
\end{eqnarray}
Using the method of lines for the temporal discretization gives us
considerable freedom in choosing numerical methods for the spatial 
derivatives, as well as the ability to choose methods of 
arbitrary orders of accuracy.  This freedom allows us to naturally
and consistently combine 
both the CENO and SBP methods in the GRHD code.

\subsection{Einstein equations}

As described in~\cite{Palenzuela:2006wp} our implementation of the 
Einstein equations takes advantage of several techniques tailored
to the symmetric hyperbolic properties of the generalized 
harmonic formulation we use.  At the linear level, these techniques 
guarantee that the full AMR implementation is stable.  
We use second and fourth order spatial derivative operators which satisfy 
summation by parts.  These operators allow one to obtain a semi-discrete
energy estimate which, together with suitable boundary conditions and
time integration, ensure the stability of the implementation of linear
systems (see~\cite{GKO}, also~\cite{Calabrese:2003vx} and references cited 
therein).  Relatedly, we employ a Kreiss-Oliger dissipation operator which is 
consistent with the summation by parts property.  

For the outer boundaries, we implement Sommerfeld boundary conditions and
follow the prescription given in~\cite{RLS07}.  
We have also used maximally dissipative boundary conditions,  
but found that they led to larger reflections at the boundaries which, 
in turn, corrupt the waveform extraction at late times.

We set the constraint damping parameters to
$\sigma_0=\sigma_1=1$.  These values were previously used in both
binary black hole and boson star evolutions, and work similarly in
the binary neutron star evolutions presented here.  For the cases
discussed here, constraint violations remain under control during
the evolutions.

Finally, while our GH formalism allows for general coordinate 
choices through the source functions $H^a(t,x^i)$, 
we adopt $H^a(t,x^i)=0$ in all the  
simulations described here. Thus, the coordinates adopted are strict 
harmonic coordinates.

\subsection{Perfect fluid equations}

The perfect fluid equations are integrated using an HRSC solver based
on the CENO method~\cite{LiuOsher}, incorporating some modifications by
Del Zanna and Bucciantini~\cite{DelZanna:2002qr},
Detailed discussions of our method 
have been presented previously~\cite{Neilsen2005,Anderson:2006ay}.

We choose the CENO method to solve the fluid equations primarily for
two reasons.  
This means that the discrete fluid solution corresponds to point values
of the solution and not cell averages.
First, it is a finite difference or vertex centered scheme.
As the Einstein equations are discretized with finite differences, 
coupling these equations to the fluid equations with AMR is simplified if
both sets of variables are defined at the same grid locations.
Secondly, CENO uses
a component-wise decomposition (central schemes) of the equations rather 
than a spectral decomposition (upwind schemes).  Central schemes, are more
efficient than spectral decomposition schemes. Although they
are more diffusive at discontinuities, their solutions often differ 
only slightly
from those obtained using upwind methods.  With AMR we can sharply resolve
all interesting features of the solution.   Outflow boundary conditions are
applied at the physical outer boundary.

The HLLE flux is used for the numerical flux~\cite{Harten}.
This is a central-upwind
method that uses the largest eigenvalues of the Jacobian matrix 
in each direction.
To calculate the numerical fluxes, we choose to use piecewise parabolic 
method (PPM) reconstruction for the 
fluid variables~\cite{Colella:1982ee}, and reconstruct the primitive
variables.  No dissipation or discontinuity detection is used in the
reconstruction.
This is a bit of a departure from the CENO scheme.  
In general, ENO methods use a hierarchical reconstruction, where, 
for example,
a second-order reconstruction depends on an underlying first order 
reconstruction.
We have found, at least for the resolutions considered here,
 that 
CENO often favors a first order reconstruction at the center 
of stars, because of the manner in which candidate second order
stencils are compared for their similarity to the first order reconstruction.
This loss of
accuracy at the center of the star damps the physical quasi-normal 
oscillations of the star, and can lead to a long-term growth of the central
density.  PPM gives a superior reconstruction for stellar interiors, and
therefore we adopt this reconstruction here.
When the fluid flow is highly relativistic, the reconstruction procedure
can produce unphysical states.  When this occurs, we attempt reconstruction
using a lower order.  For example, if PPM fails, then a linear
minmod reconstruction is attempted, and if this fails, then no reconstruction
is used.

A consequence of using HRSC methods is the need to go back and forth 
between primitive, $\bbw$, and conservative, $\bbq$, variables.  
While the relation of the conservative variables in terms of the primitive
variables is algebraic, the transformation that gives the primitive variables 
in terms of the conservative variables 
is transcendental.  We use a Newton-Raphson solver designed for the
MHD equations to find the primitive variables~\cite{Neilsen2005}.  
At grid points where this solver may fail, 
the primitive variables are obtained from neighboring points by linear 
interpolation.
The conservative variables are then recalculated at these points from the 
interpolated primitive variables.

Unphysical states can arise during the evolution of the fluid
equations.  This often occurs in evacuated regions of the grid, where
truncation errors or effects from finite precision arithmetic are 
significant compared to the fluid densities.
To compensate for some  of these errors, a floor is applied to
the energy variables $\tilde D$ and $\tilde \tau$ as
\begin{eqnarray}
\tilde D &\leftarrow&  \max(\tilde D,\, {\rm floor}),\\
\tilde \tau &\leftarrow&  \max(\tilde \tau, \,{\rm floor}).
\end{eqnarray}
The floor in these runs is set
between $1\times 10^{-8}$ and $5\times 10^{-9}$, which is seven
orders of magnitude smaller than
the central rest mass densities ($\rho_c$) of the individual stars.  
The floor value must be small compared to the densities in the problem
so that the floor does not significantly affect the dynamics of interest.
Often the effect of the floor can only be ascertained by varying
it in a series of runs.  For example, we found that floor values
of $10^{-7}$ are too large, producing a noticeable increase in $\rho_c$
during the evolutions, and changes in the stellar trajectories and
the emitted waveforms.  These errors essentially disappear when the floor is
$10^{-8}$, and reducing it further to $5\times 10^{-9}$ does not change
the solutions.  Thus, we adopt here a floor of $1\times 10^{-8}$.

%
%
\section{Oscillating modes of single TOV stars}

As a first test of our combined GRHD code we consider a single TOV star.  Our
goal is not only to represent the analytic TOV solution, but to
accurately reproduce the known radial oscillation modes of the star.
While the TOV solution is spherically symmetric and static,
discretization effects act as small perturbations that excite the
normal modes of the star.

\begin{figure}
\begin{center} 
\epsfig{file=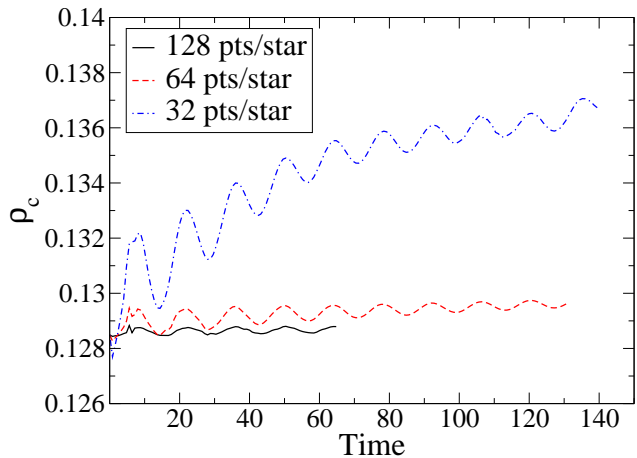,height=9.5cm,angle=270}
\caption{This figure shows oscillations in the central rest energy
density, $\rho_c$, for a dynamic spacetime evolution
of a single TOV star at three different resolutions:
32, 64, and 128 points across the star. 
The initial data are for a star of mass
$M = 0.14$, circumferential radius $0.958$, central rest mass
density $\rho_c = 1.28 \times 10^{-1}$, $\Gamma=2$, and $\kappa = 1$.
The outer boundary of the simulations is 12 stellar radii away from the 
center of the star, and PPM is used to reconstruct the fluid variables.
While $\rho_c$ increases noticeably for the coarsest resolution run, 
it eventually stabilizes at a higher value, giving a stable configuration.
Results from the Fourier transform of this data are given in 
Table~\ref{table:frequencies}.
Owing to the computational costs of these simulations, the higher resolution
runs were not evolved to the same end time.  
In particular, the highest resolution run was evolved only until
$t\simeq 65$.
}
\label{fig:single_tov}
\end{center} 
\end{figure}

The initial data for this test consist of a $\Gamma=2$ polytrope with
$\kappa=1$.  (The solution is calculated using a modified version of the
RNS code of
Stergioulas~\cite{RNS}.)  
The star, in the geometrized units with
$\kappa = 1$, has a mass of $M = 0.14$, a circumferential radius $R=0.958$,
and central rest mass  density $\rho_c = 1.28 \times 10^{-1}$.
We evolve the data in a dynamic spacetime at different resolutions
and using different reconstruction methods for the fluid variables.  
Figure~\ref{fig:single_tov} shows $\rho_c$ plotted as a function of time 
for three resolutions of
32, 64, 128 points across the star.  As expected, the oscillations and overall
drift in $\rho_c$ converge with resolution.  This is important both as
a code test and an indication of the resolution necessary to capture some
dynamics of stellar interiors.  The data in Figure~\ref{fig:single_tov}
were generated using PPM reconstruction.  We found that first- and 
second-order CENO reconstructions were more diffusive, 
resulting in larger drifts in $\rho_c$.
Consequently, we had difficulty in reproducing the radial pulsation
modes of the star using these reconstructions.

To confirm that the code reproduces the expected physical behavior, we
examine the radial pulsations of the star.  The modes are calculated from
the oscillations in $\rho_c$, and the extracted frequencies are
shown in Table~\ref{table:frequencies}. 
(Though we present data for the central density only here,
we have verified that these are global modes by examining the
time variation of density and velocity in the star.)
 These oscillation modes can
be compared to the known radial perturbation modes~\cite{Font2002},
and these frequencies are in excellent agreement.  
Note, to make these comparisons we rescale the perturbation results
as described in the Appendix,
which were calculated for 
$\kappa=100$.
These validations are
a stringent test of our computational methods and give us considerable
confidence that our code accurately reproduces the physics of these
systems.

\begin{table}
\begin{tabular}{c|c|c|c}
\hline
Mode & 3D GRHD code & Perturbation results & Relative Difference \\
     & (kHz) &  (kHz) & (\%) \\
\hline
F  &  14.01    & 14.42 & 2.88 \\
H1 & 39.59     & 39.55 & 0.1  \\
H2 & 59.89     & 59.16 & 1.2  \\
H3 & 76.94     & 77.76 & 1.1  \\
\end{tabular}
\caption{Comparison of small radial pulsation frequencies for an
  evolved star using the 3D GRHD code to the linear perturbation 
  modes~\cite{Font2002}.  The polytrope is constructed for $\Gamma=2$ and
$\kappa =1$.  The perturbation
  results have been appropriately rescaled for $\kappa = 1$~\cite{Noblephd}.
  The Fourier transform of the central density time series is plotted in 
  Figure~\ref{fig:rho_fft}.
  }
\label{table:frequencies}
\end{table}

\begin{figure}
\begin{center} 
\epsfig{file=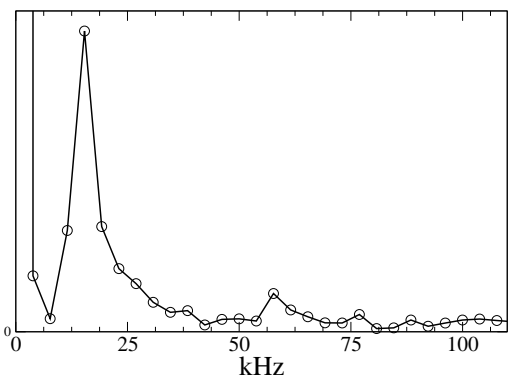,height=8.5cm,angle=270}
\caption{This figure shows the Fourier transform of the oscillations of $\rho_c$ seen
  in the highest resolution simulation of Figure \ref{fig:single_tov}.  
    Five distinct peaks are observed; the first four peaks
    are compared with results found via linear perturbation 
    (See Table \ref{table:frequencies}).  The scale of the vertical axis is arbitrary.} 
\label{fig:rho_fft}
\end{center} 
\end{figure} 

%
%
\section{Binary neutron stars}

In this paper, we consider two
different binary neutron star mergers, 
one resulting in a prompt collapse to a black hole and one that
results in a differentially rotating neutron star which persists for
a long time (as compared to an orbital time close to merger).
We evolve the systems through several orbits and extract 
gravitational radiation from the orbiting phase and the merger.
In the course of performing these
evolutions, we carefully examine some numerical questions to ensure the
accuracy of our results.

Initial data for both binaries are set by superposing the initial data for
single, boosted TOV stars~\cite{Matzner:1998pt}.
Provided that the initial separation
between stars is sufficiently large, violations in the momentum and
Hamiltonian constraints are at or below the truncation error threshold.
We monitor that this is indeed the case for our chosen separations by
evaluating the constraints and checking that any violations are of the
same order as those obtained for the single stars considered in the previous
section. Thus, these data are numerically consistent.
The boost velocities are smaller than the 
corresponding Keplerian velocities for Newtonian circular orbits.
Thus, our data are not quasi-circular 
(as used in~\cite{Miller:2003vc,Shibata:2002jb}), and they do not correspond
to a system resulting from a long, slow inspiral.
However, these data allow us to both test our implementation, as well
as to examine how radiative effects circularize the orbits.
Forthcoming work will consider initial data taken from post-Newtonian and
quasi-equilibrium approaches.

We extract the gravitational wave information by
computing the Weyl scalar $\Psi_4$, and for convenience we further
decompose $r\Psi_4$ as an expansion in terms of
(spin-weighted) spherical harmonics
\begin{eqnarray}
r \Psi_4 = \sum_{l,m} C_{l,m}\, {}^{-2} Y_{lm}.
\end{eqnarray}
This extraction is done at three different 
locations from the center of mass, and we shift the obtained
quantities in time to
account for the travel time between the observers along null rays.
These observers are placed within the wave zone, and the shift in time 
is given simply by the distance in Minkowski spacetime between the observers.
As we consider here only equal mass binaries, corrections for gauge effects
should be small~\cite{lehnermoreschi}.  An analysis
of these effects for different binaries will be presented 
elsewhere~\cite{gaugeradiation_us}.

\begin{figure}
\begin{center} 
\epsfig{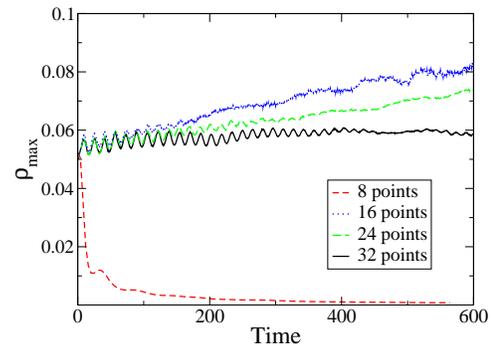}
\caption{This figure shows the maximum value of $\rho_o$ in binary 
simulations at four resolutions: 8, 16, 24 and 32 points across each star. 
With fewer than 16 points across the star, the stars disperse.
For increasing resolutions, the solutions converge.}
\label{fig:rho_convergence}
\end{center} 
\end{figure} 

\begin{figure}
\begin{center} 
\epsfig{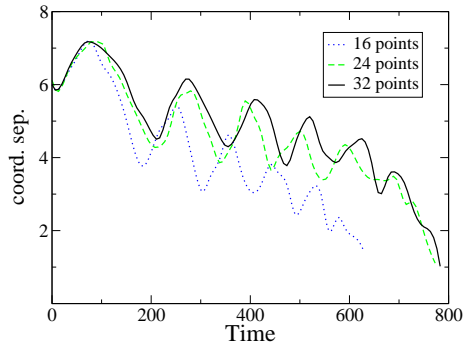}
\caption{The coordinate separation between the stars in a merging binary is
shown here as a function of time for three resolutions. 
Notice that the merger time, about $t=800$, is almost the same
for the two finer resolutions.}
\label{fig:rad_convergence}
\end{center} 
\end{figure} 

As discussed in~\cite{Miller:2003vc}, boundary and resolution effects
can strongly influence the dynamics of these systems.
To explore the effects of outer boundaries on the simulation results,
we perform two otherwise identical evolutions with outer
boundaries at different locations.  While a more detailed discussion
of these tests follows below in Section~\ref{sec:binary_black_hole}, 
we find that outer boundaries at 80 stellar radii have negligible influence
on the solution.
To examine resolution effects, we adopt 
a threshold error tolerance for the shadow hierarchy such that
the resulting mesh covers each star with a minimum of 16 points.
While in the previous section we used much higher resolutions to capture
the interior dynamics of single stars, binary evolutions
at similar resolutions here
are prohibitively expensive.  Figure~\ref{fig:rho_convergence}
gives an indication of the minimum resolution required to 
evolve the binary without resolving the internal dynamics of individual
stars.  Figure~\ref{fig:rad_convergence} shows the (coordinate) radial
distance between the center of the stars versus time for the three different 
resolutions.  The trajectories converge as the resolution is
increased. 

\begin{figure}
\begin{center} 
\epsfig{file=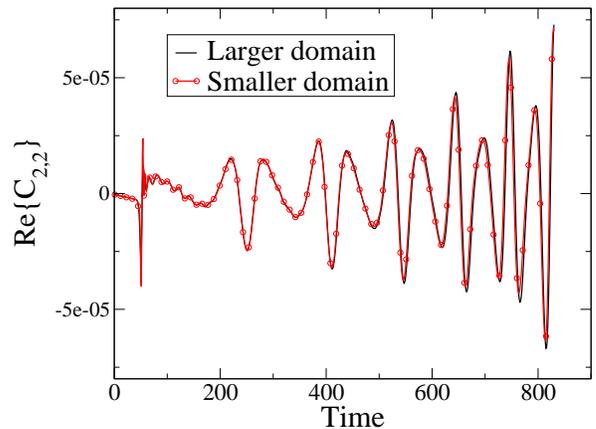,height=8cm,angle=270}
\epsfig{file=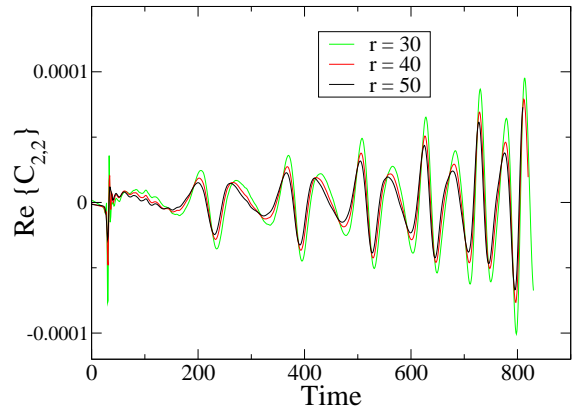,height=8.0cm,angle=270}
\caption{{\it Top Panel}. The $\ell=2$, $m=2$  mode of $r\Psi_4$
  extracted at 50 stellar radii for
  binary simulations with different domain sizes.  The smaller
  domain is of size $\pm 80$ stellar radii while the larger is 
  $\pm 124$ stellar radii.  The two results differ only by a small 
  phase and amplitude error which appears late in the evolution.  
  For both simulations, the floor value is $1\times 10^{-8}$.
 {\it Bottom Panel}:  This includes three plots of the 
$\ell=2$, $m=2$ mode of $r\,\Psi_4$ extracted at 30, 40, and 50 stellar radii.
The initial data are described in Section~\ref{sec:binary_black_hole}. 
The domain of the calculation is 248 stellar radii 
across. The signals from different extraction surfaces are shifted in time 
by the appropriate (flat-space) differences between the extraction radii. 
} 
\label{fig:boundary}
\end{center} 
\end{figure}

\subsection{Black hole final state}
\label{sec:binary_black_hole}

\begin{figure}
\begin{center} 
\epsfig{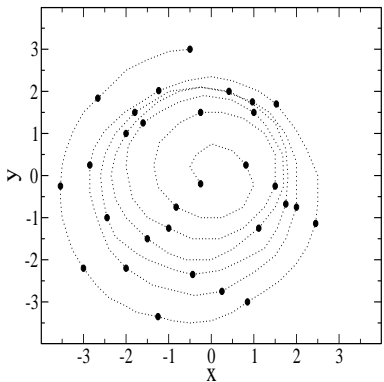}
\caption{The coordinate trajectory of the center of one of the neutron stars 
as it spirals into a black hole end state.
The points (filled circles) that have been included along the trajectory 
are the coordinate locations of the maximum density. These points are shown
at intervals of 
$\Delta t=20$ in order to give an idea of the star's speed.}
\label{fig:coord_trajectory}
\end{center} 
\end{figure}

\begin{figure}
\begin{center} 
\epsfig{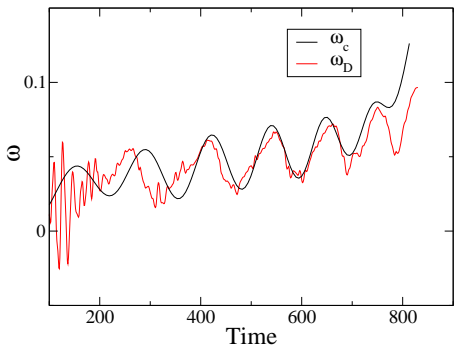}
\caption{Orbital frequency of the binary as calculated 
from the numerical evolution in  
two different ways. $\omega_c$ is obtained by following the coordinate 
position of the centers of the stars while $\omega_D$ is obtained from the 
dominant mode of $r\Psi_4$.}
\label{fig:exc_omega}
\end{center} 
\end{figure}

\begin{figure}
\begin{center} 
\epsfig{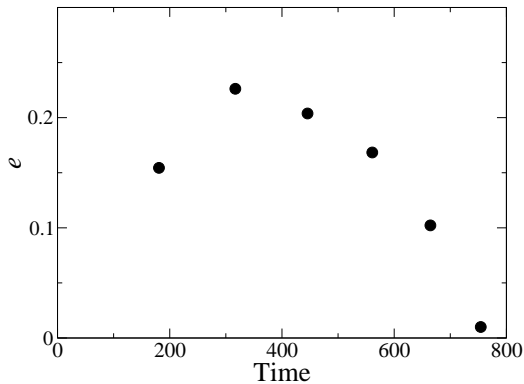}
\caption{The eccentricity obtained from Eq.~(\ref{eq:eccentricity}). After a 
transient behavior due to the initial configuration, an overall monotonically
decreasing behavior is seen in the eccentricity as the binary
orbit becomes tighter.} 
\label{fig:eccentricity}
\end{center} 
\end{figure}

The first set of initial data gives a binary neutron star merger that
results in a prompt collapse to a black hole.  As mentioned previously, the initial
data are constructed from superposing two equal mass
neutron stars with zero 
spin angular momentum.  In particular, each star has a mass of
$M=0.89~M_{\odot}$, a radius of $R=16.26$~km, and a central density of
$3.24 \times 10^{14} \; {\rm g}/{\rm cm}^3$.  
The stars are placed initially at the coordinate locations
$(x,y,z) = (0, \pm 3,0)$ with the boost $v^i = (\mp 0.08,0,0)$.

We first investigate possible effects from the outer boundaries on
the simulation results by performing two otherwise identical
evolutions with the outer boundaries at different locations.  In one,
the outer boundary is at $80R$, and in the other it is at $124R$.
These simulations use the shadow hierarchy, and the AMR
grid-structure is determined by the threshold error parameter.
An additional set of fixed
fine grids is placed at larger distances to ensure sufficient
resolution for computing waveforms.  As a consequence,
the grid-structure in the central region is determined dynamically
while at far distances it is kept fixed.  We compare the $C_{2,2}$
component of the gravitational wave signal measured by an observer
at a fixed coordinate distance, $50R$, for the two computational
domains.  These waveforms are shown in Figure~\ref{fig:boundary}.
which shows only small differences in the waveforms at late times.
Additional tests indicate that these differences 
arise from the location of the exterior, fixed refinement
boxes. This observation is indicated by 
the coincidence of results obtained with outer boundaries at $100 R$ and
$80 R$ with exactly the same coordinate locations of the exterior
grids.  The overall excellent
agreement between the wave signals suggests that the influence of the
boundary location is negligible.

The dynamics of the subsequent evolution shows a clear eccentricity which is 
reflected both in the gravitational waveforms  
(bottom panel of Figure~\ref{fig:boundary}) and the 
coordinate trajectories (Figure~\ref{fig:coord_trajectory}).  It is worth noting that, following the 
suggestion of~\cite{BCP07}, a waveform similar to 
Figure~\ref{fig:boundary} can be obtained by using the Newtonian 
quadrupole approximation with the coordinate trajectories from 
Figure~\ref{fig:coord_trajectory}.  In addition, these trajectories 
are similar to those obtained by integrating the 2.5 post-Newtonian 
equations.
Finally, as with the black hole case reported in~\cite{BCP07},
the orbital coordinate frequency $\omega_c$ (computed from the coordinate
trajectories) is in good agreement with the orbital waveform frequency 
$\omega_D$ (computed from the dominant mode $\ell=2$, $m=2$ of $r\Psi_4$), as
shown in Figure~\ref{fig:exc_omega}.

The eccentricity can be computed using the Newtonian definition given in~\cite{MorWill02}
\begin{equation}
   \textit{e} = \frac{\sqrt{\omega_p} - \sqrt{\omega_a}}%
{\sqrt{\omega_p} + \sqrt{\omega_a}} ,
\label{eq:eccentricity}
\end{equation}
where $\omega_p$ is the orbital frequency at a local maximum and $\omega_a$ 
the subsequent local minimum. The eccentricity of this 
simulation is shown
in Figure~\ref{fig:eccentricity}. To compute this, we take each half-cycle 
and evaluate expression (\ref{eq:eccentricity}), thus obtaining a discrete 
set of values. The first point is clearly affected by the 
initial data adopted, but the subsequent points show an overall 
decrease towards zero.  This is expected as the gravitational 
radiation carries away angular momentum, and its loss 
circularizes the orbit.

\begin{figure}
\begin{center} 
\epsfig{file=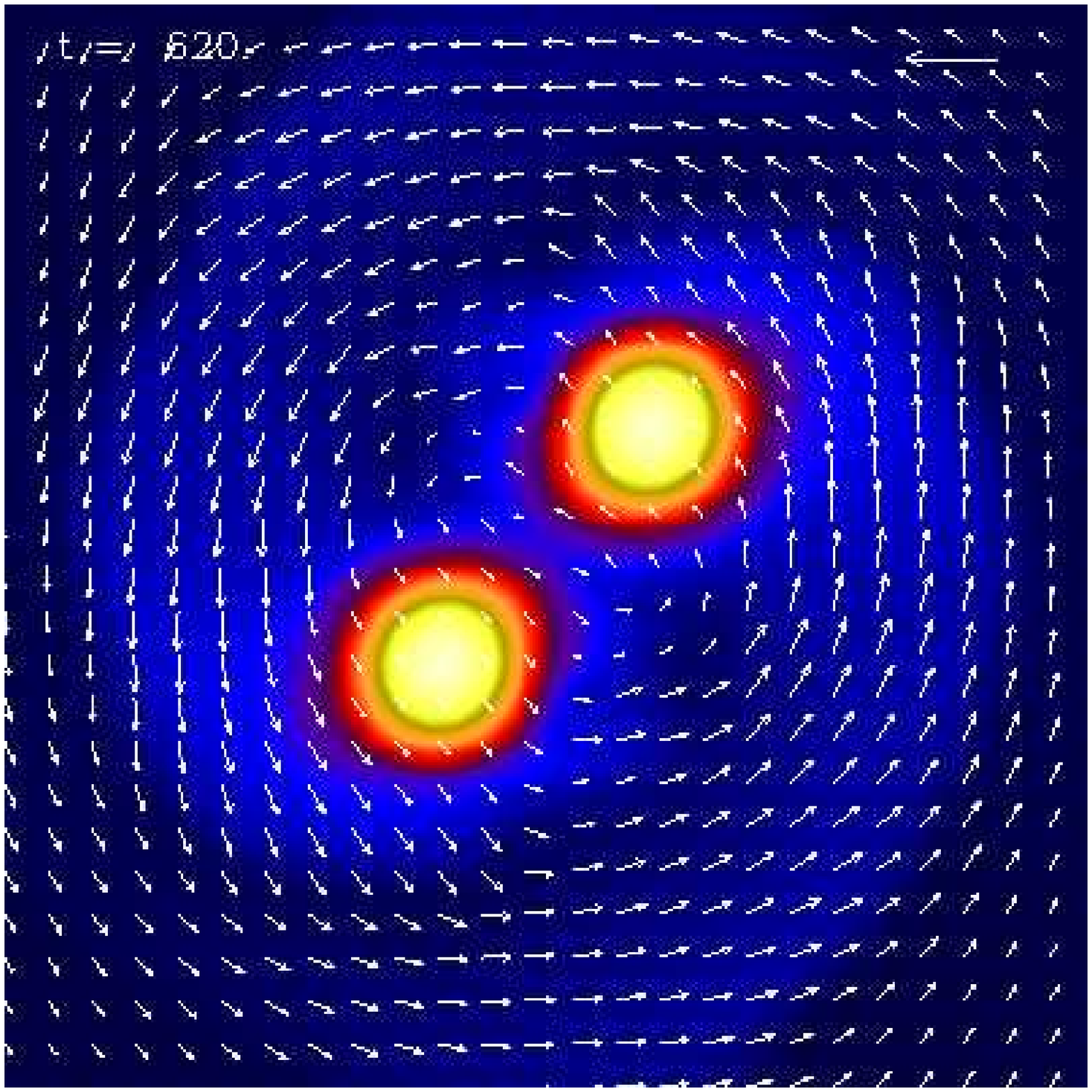,height=4.5cm}
\epsfig{file=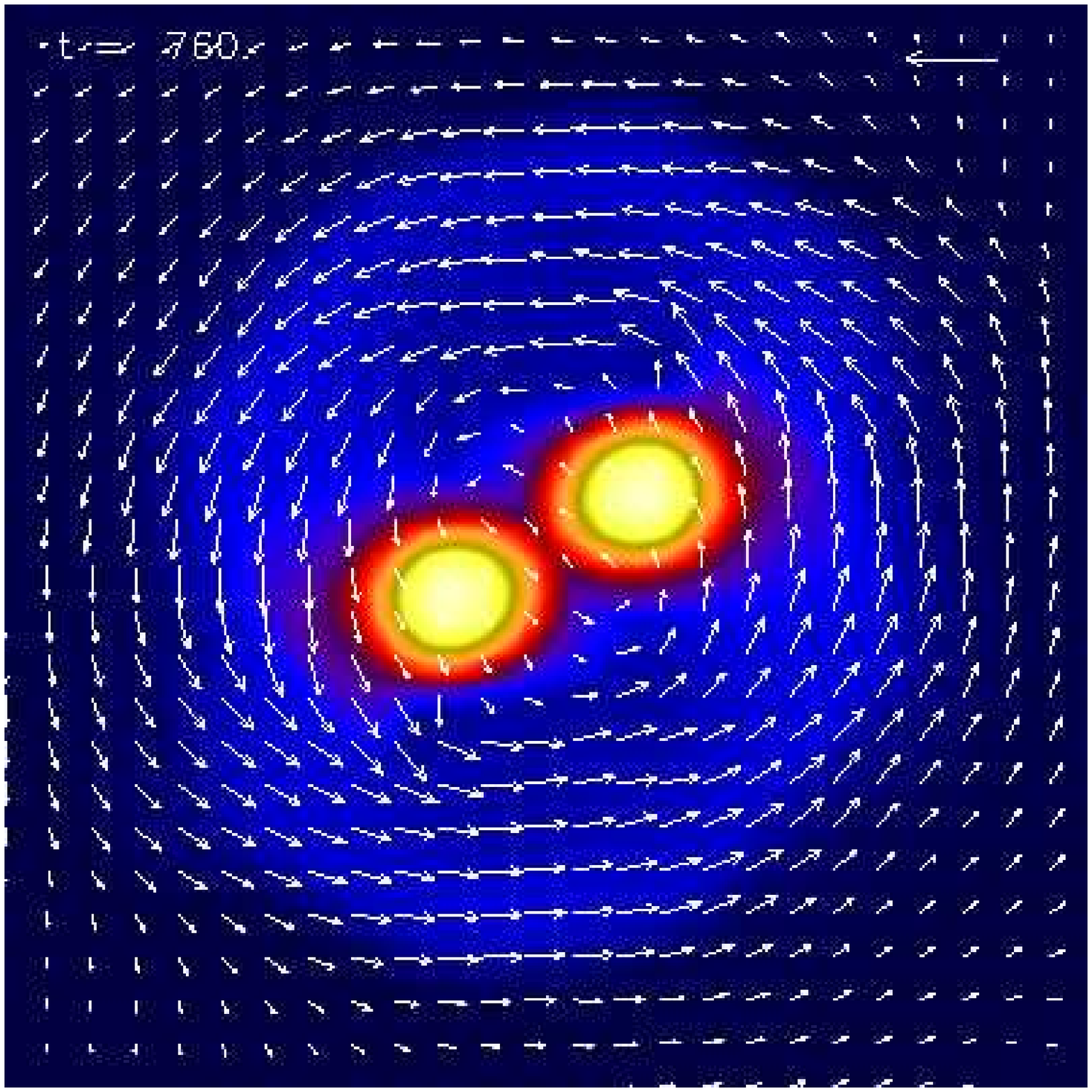,height=4.5cm}
\epsfig{file=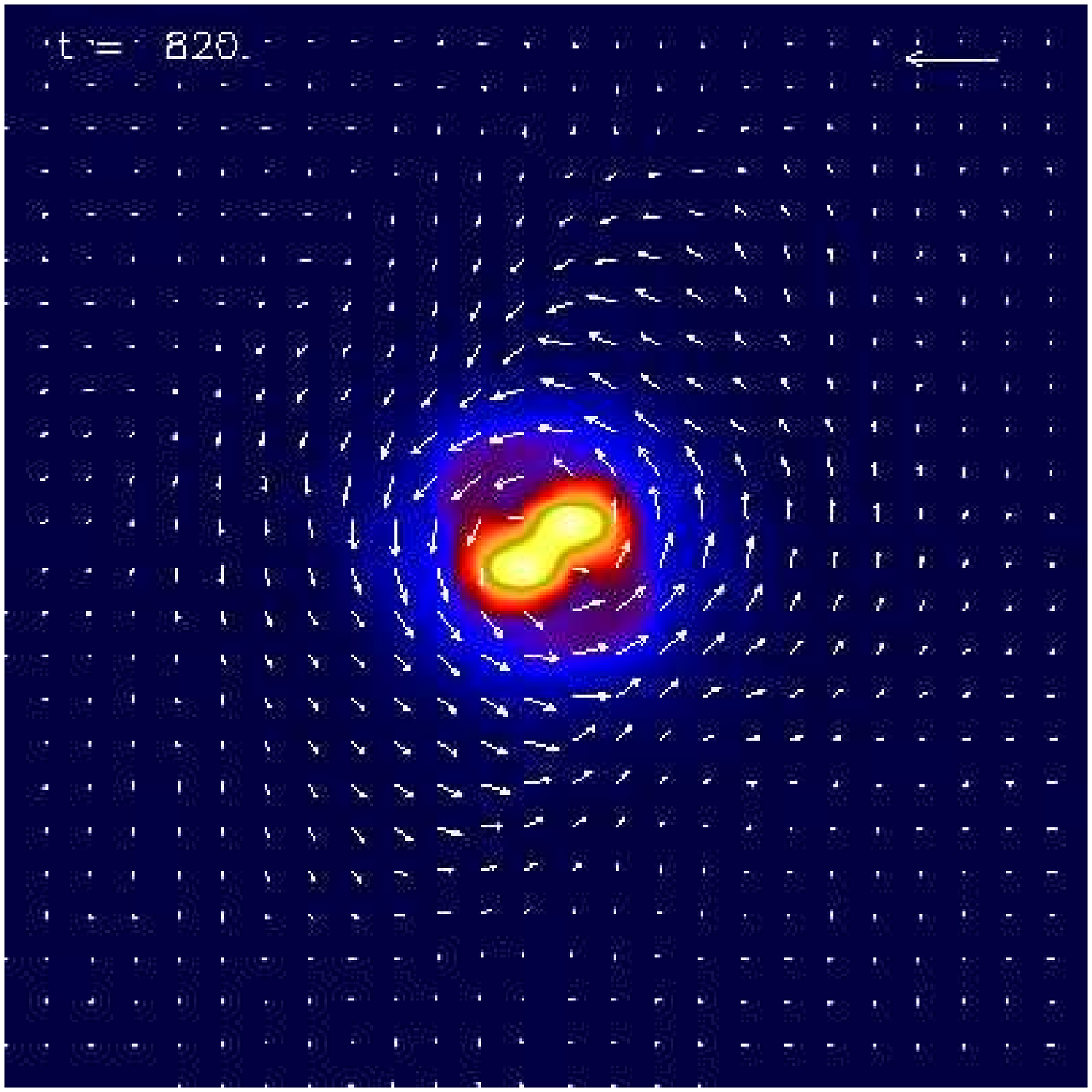,height=4.5cm}
\caption{Snapshots at select times viewed down the $z$ axis 
of the orbiting stars 
and their subsequent collapse
to a black hole. These snapshots zoom in on the central region of the 
grid and show
only a twentieth of the $z=0$ slice of the computational domain. 
The stars orbit counterclockwise 
seven times before merging and collapsing to a black hole.  
The arrows indicate the fluid velocity.
The reference vector in the upper right hand corner of each panel
has a magnitude of 0.5.
The color scheme indicates the rest mass density.
The plots show the simulation at times 620, 760, and 820 as indicated
in the upper left corner of each image.
See Figure~\ref{fig:minlapse} for a plot of the
lapse at the origin as a function of time for this system.}
\label{fig:fluidsnapshots}
\end{center} 
\end{figure}

\begin{figure}
\begin{center} 
\epsfig{file=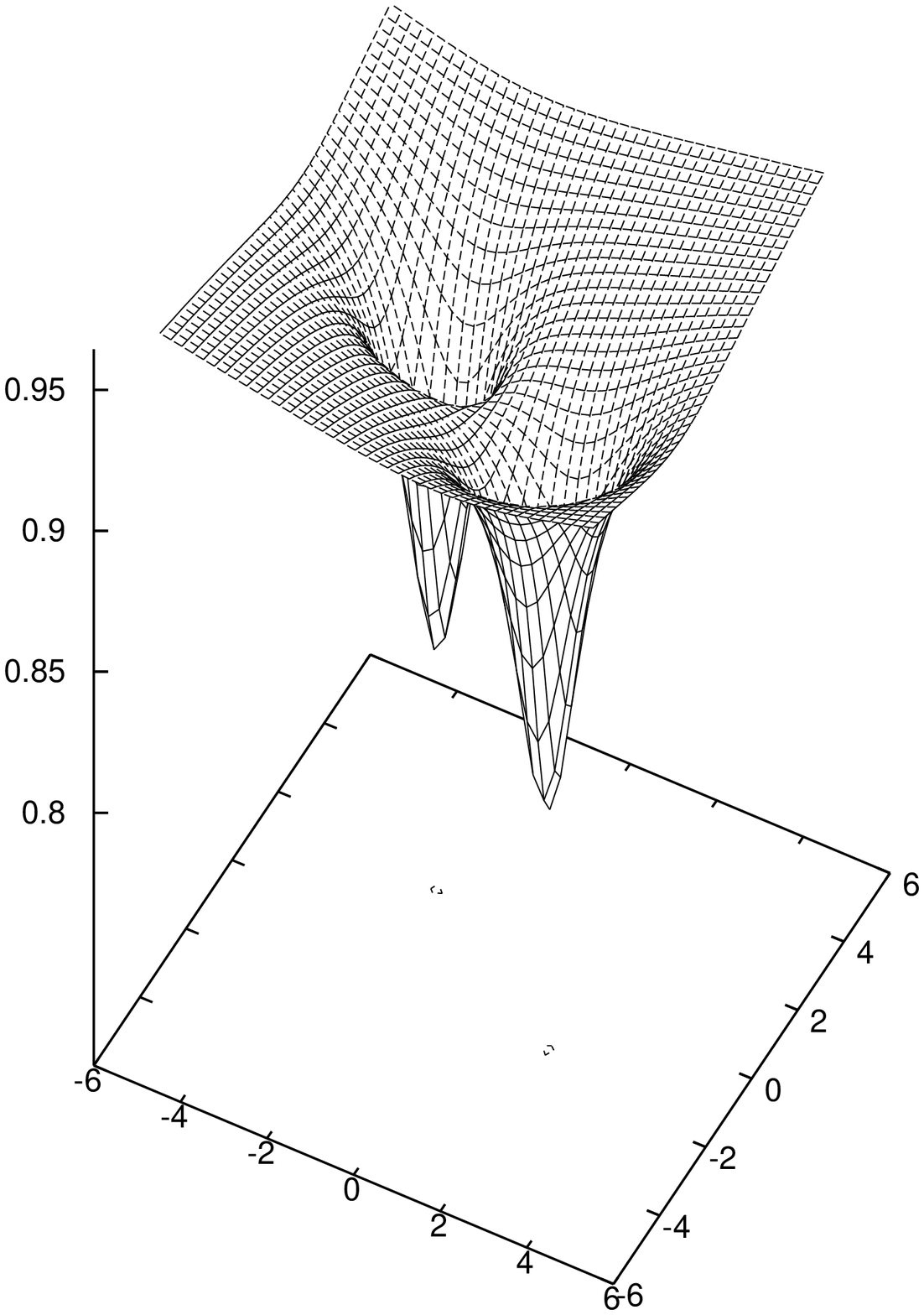,height=6.0cm}
\epsfig{file=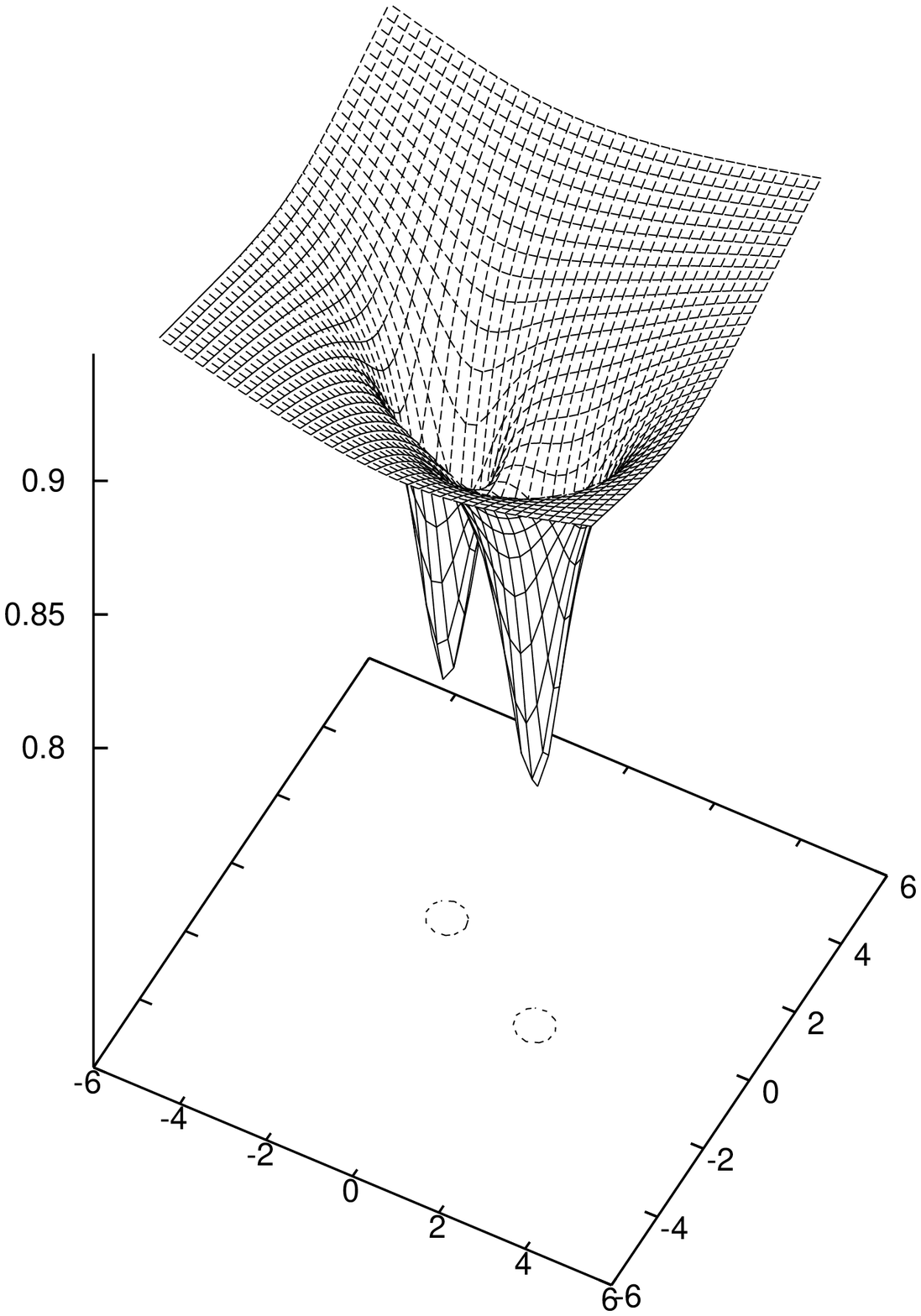,height=6.0cm}
\epsfig{file=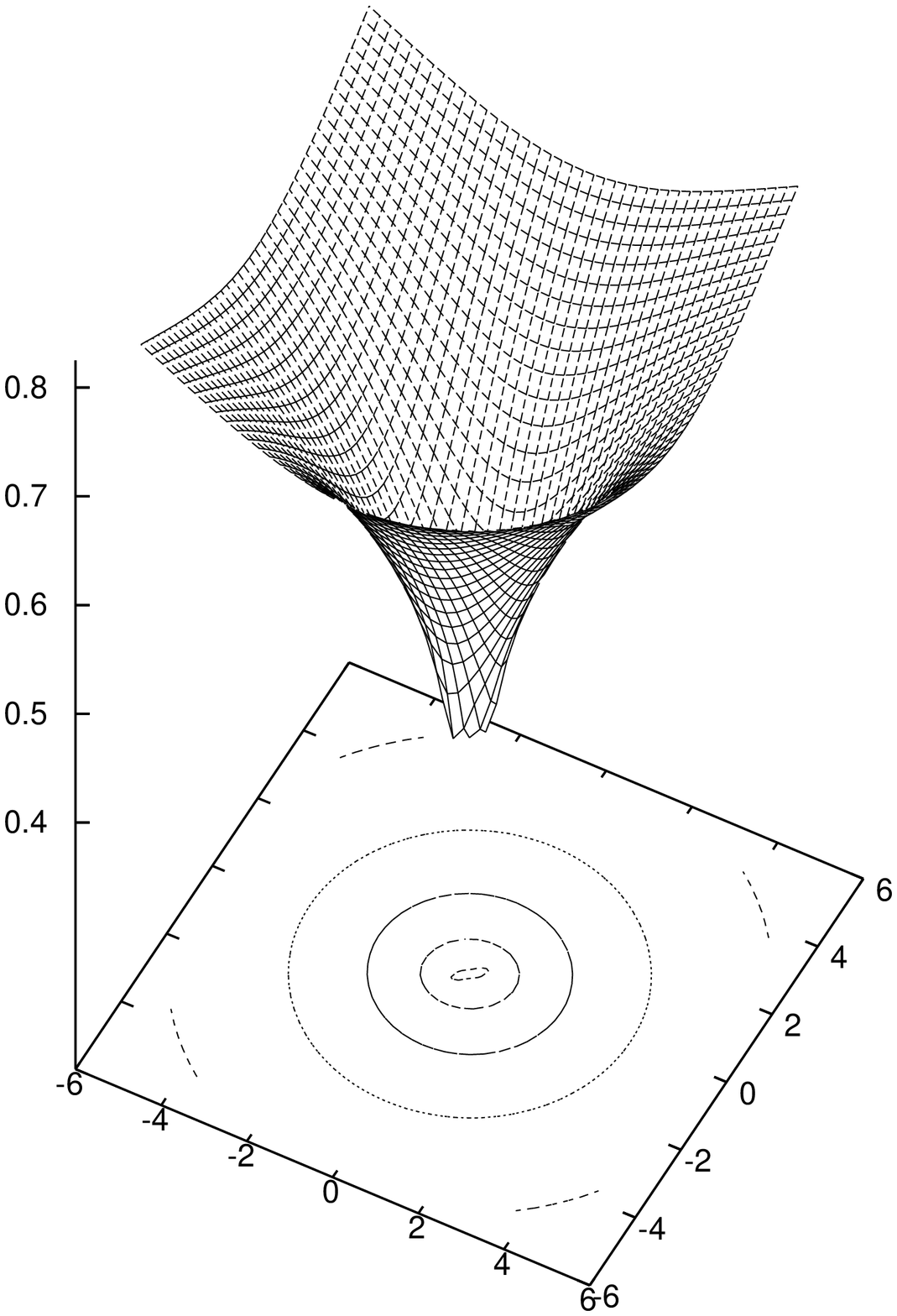,height=6.0cm}
\caption{Snapshots of the lapse on the $z=0$ plane at times 400, 600, and 820 for the system presented in 
Figure~\ref{fig:fluidsnapshots}.  The contours shown correspond to 
$\alpha=0.8,0.7,0.6,0.5,0.4$, from the outermost to the innermost one.  
At times prior to merger, only the first contour value exists. 
After merger, 
the lapse collapses, indicating the formation of a black hole. 
Notice the essentially circular shape of all the contours 
except for the innermost one at the latest time.
The lapse at the origin as a function of time is shown in Figure~\ref{fig:minlapse}.}
  \label{fig:lapsesnapshots}
\end{center} 
\end{figure}

\begin{figure}
\begin{center} 
\epsfig{file=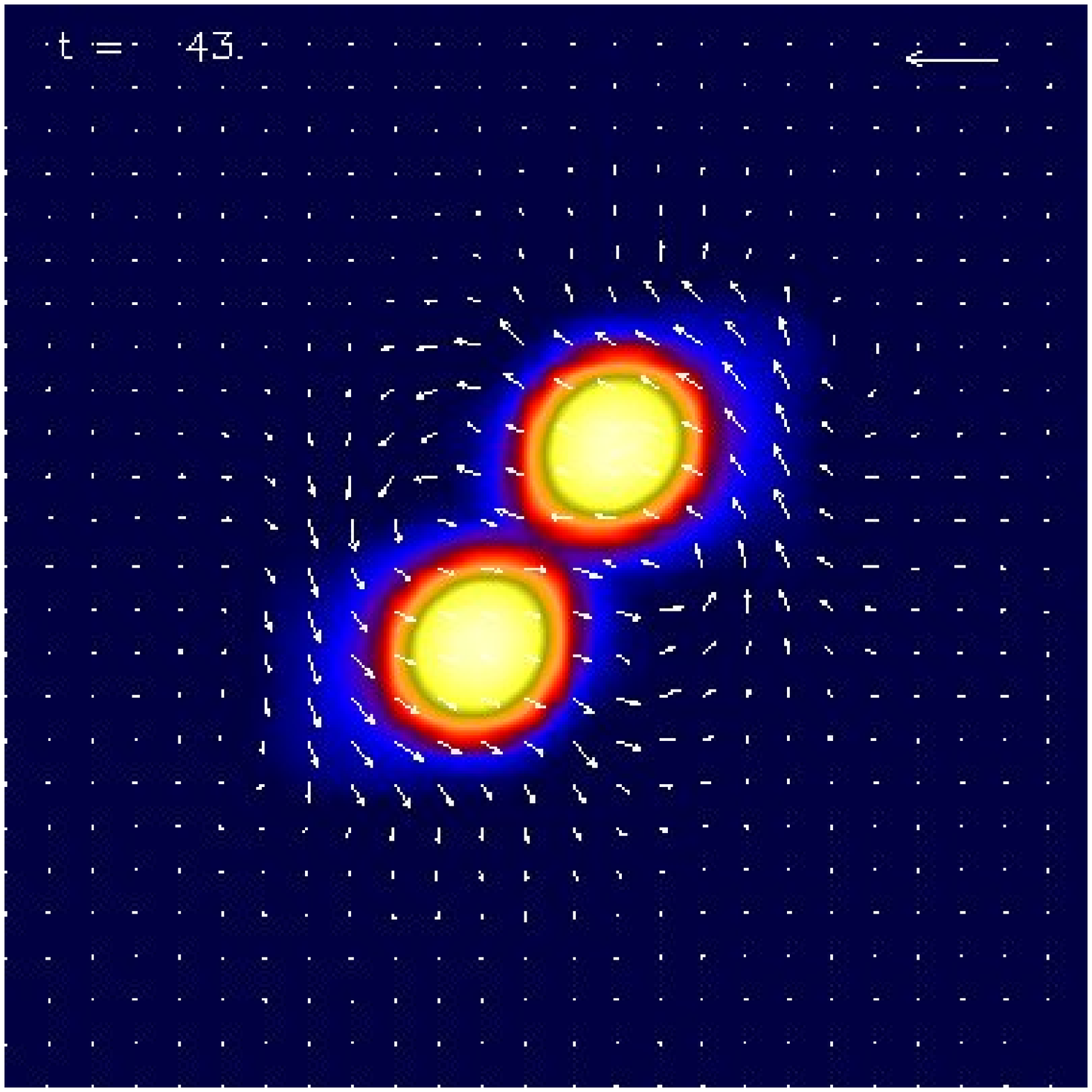,height=4.5cm}
\epsfig{file=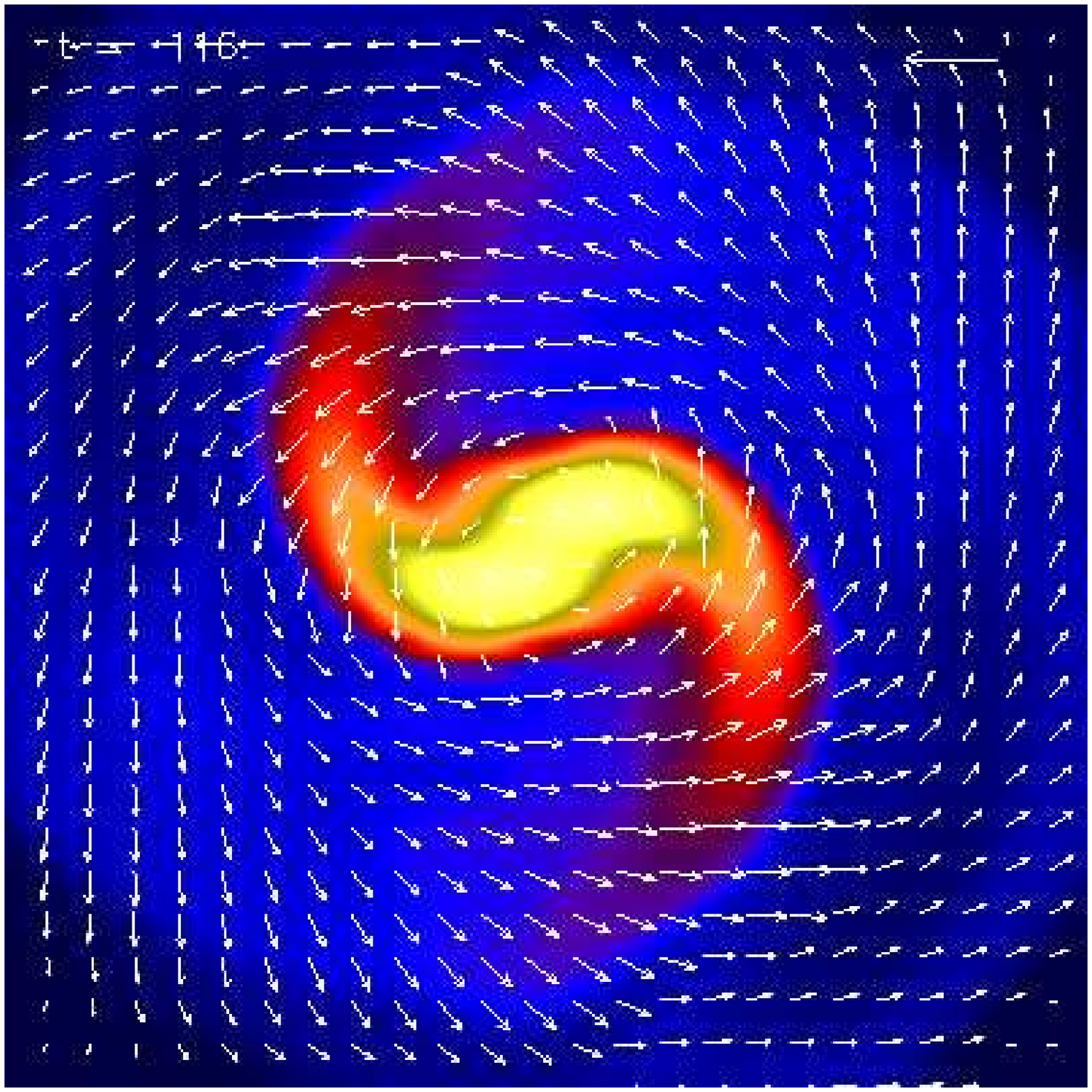,height=4.5cm}
\epsfig{file=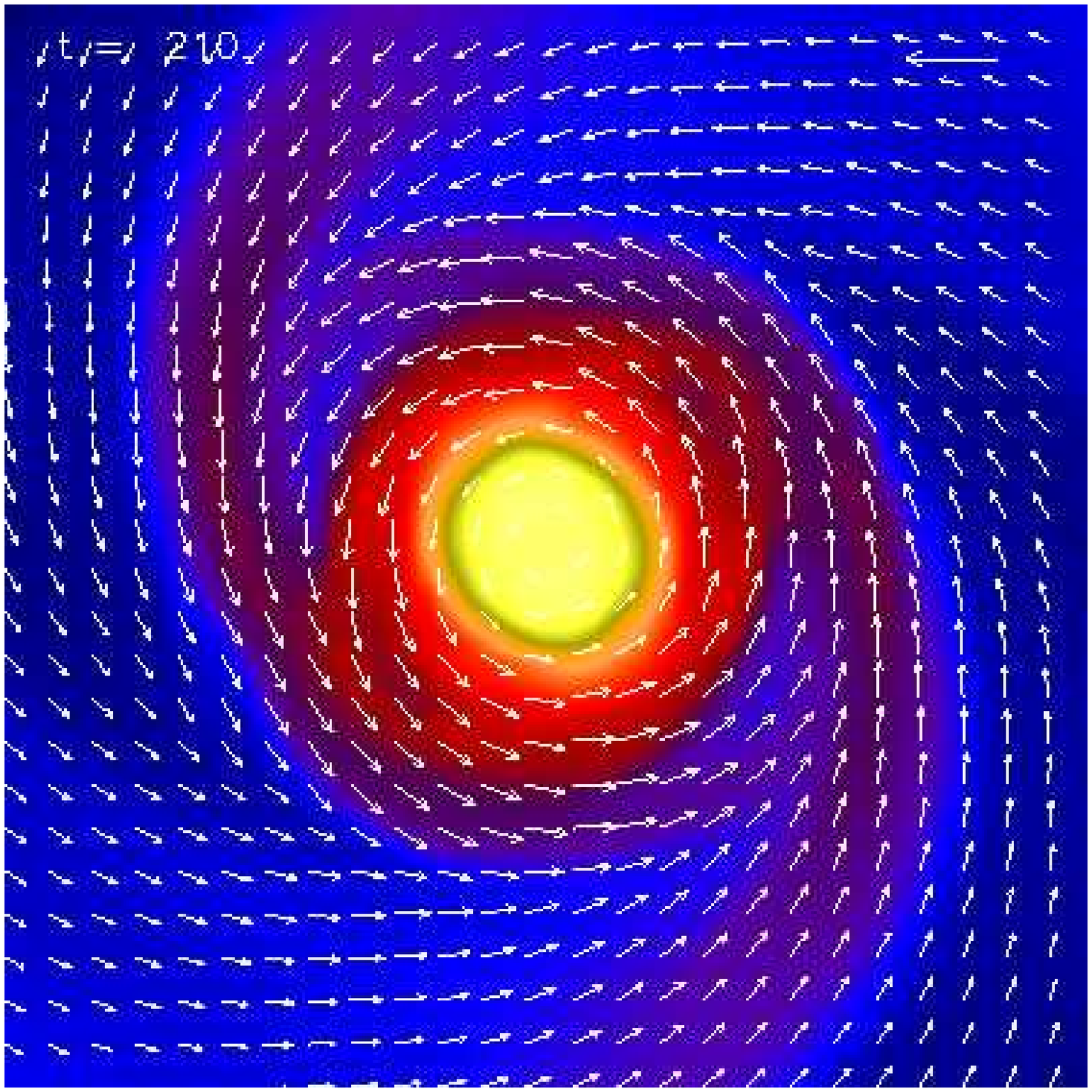,height=4.5cm}
\caption{Snapshots at select times viewed down the $z$ axis 
of the orbiting stars 
and their subsequent merger into a differentially rotating star. 
These snapshots zoom in on the central region of the  grid and show
only a twentieth of the $z=0$ slice of the computational domain. 
The stars orbit counterclockwise 
a couple of times before merging.  
The arrows indicate the fluid velocity.
The reference vector in the upper right hand corner of each panel
has a magnitude of 0.5.
The color scheme indicates the rest mass density.
The plots show the simulation at times 43, 116, and 210 as indicated
in the upper left corner of each image.
See Figure~\ref{fig:minlapse} for a plot of 
the lapse at the origin as a function of time for this system.}
\label{fig:nobhfluidsnapshots}
\end{center} 
\end{figure}

\begin{figure}
\begin{center} 
\epsfig{file=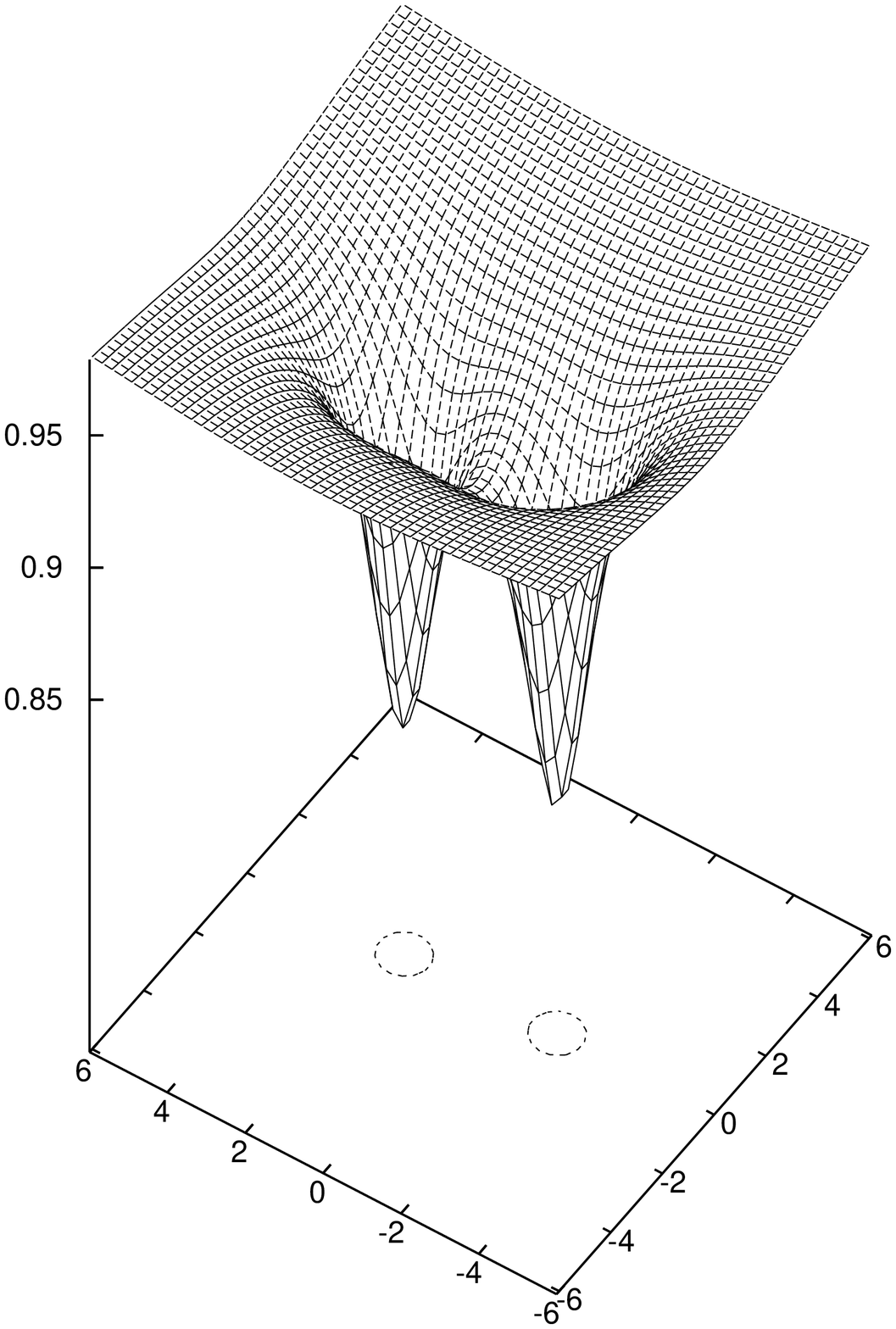,height=6cm}
\epsfig{file=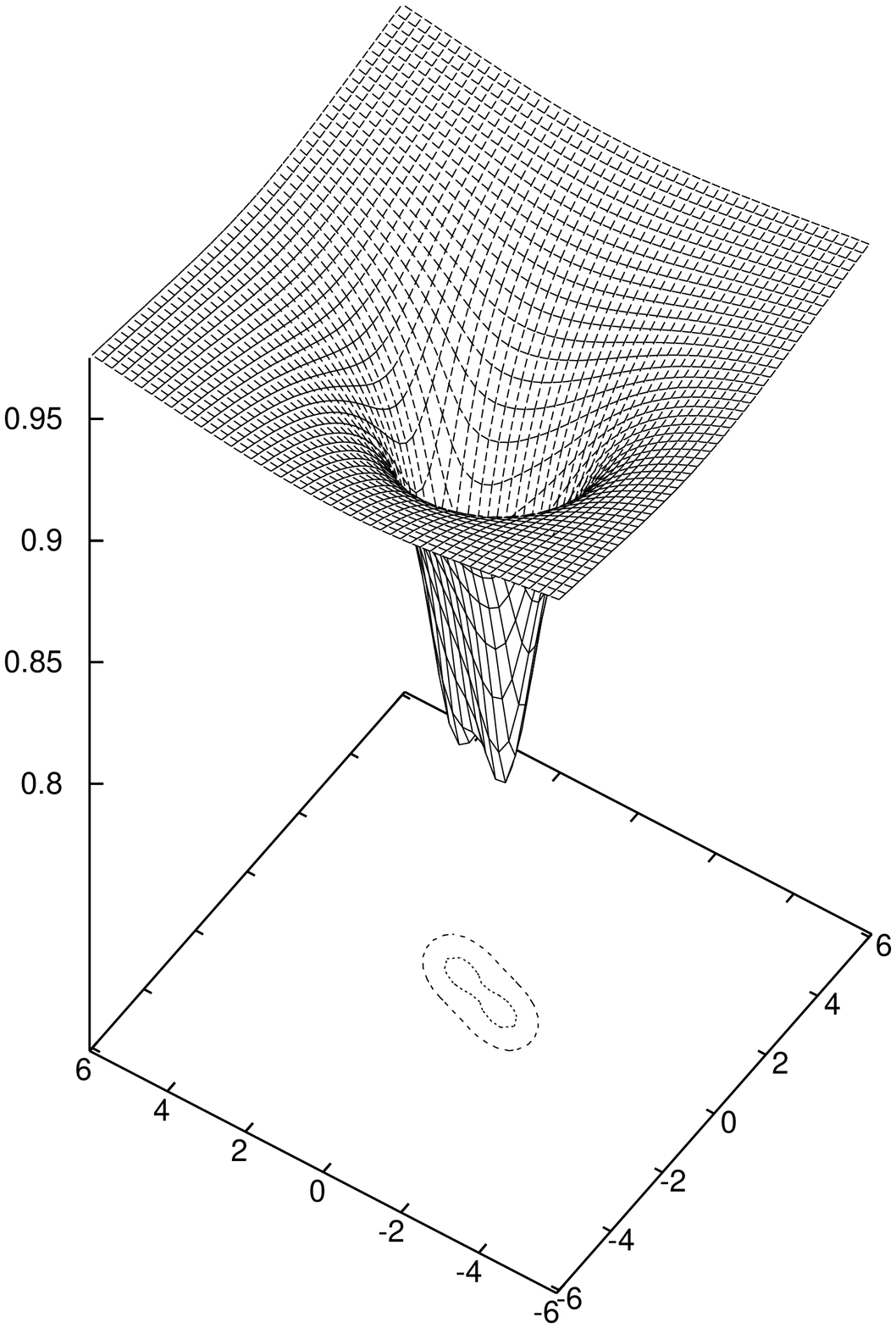,height=6cm}
\epsfig{file=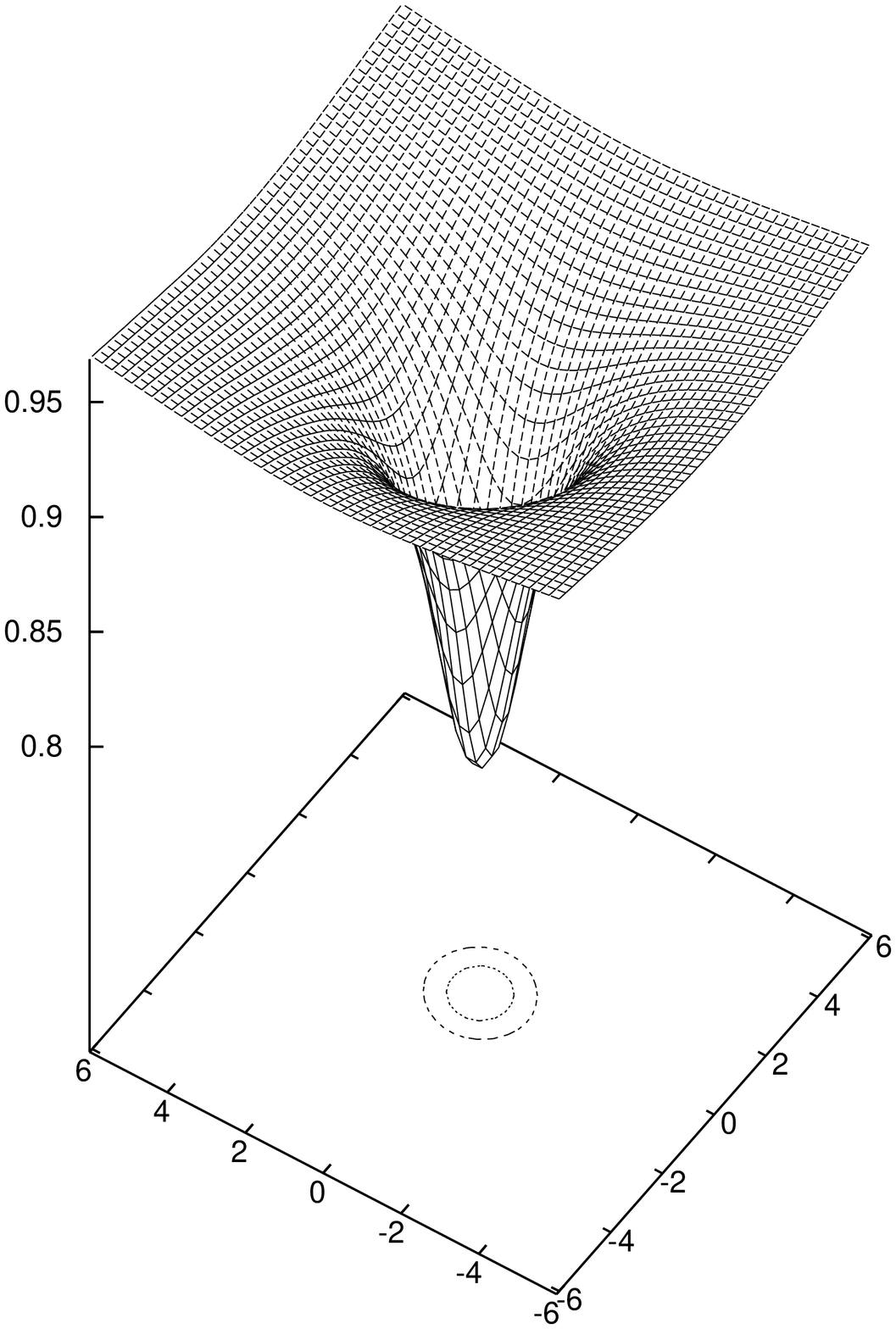,height=6cm}
\caption{Snapshots of the lapse on the $z=0$ plane at times 0, 100, and 200 for the 
system presented in 
Figure~\ref{fig:nobhfluidsnapshots}.  The contours shown correspond to 
$\alpha=0.9,0.85$ from the outermost to the innermost one.  
At early times, the contour for the lowest value is not present. After merger, 
though the lapse evolves to a slightly lower value, it remains bounded
above $\simeq 0.75$.  Notice the essentially circular shape of all the contours 
at the latest time.
The lapse at the origin as a function of time is shown in Figure~\ref{fig:minlapse}.}
  \label{fig:NOBHlapsesnapshots}
\end{center} 
\end{figure}

\begin{figure}
\begin{center} 
\epsfig{file=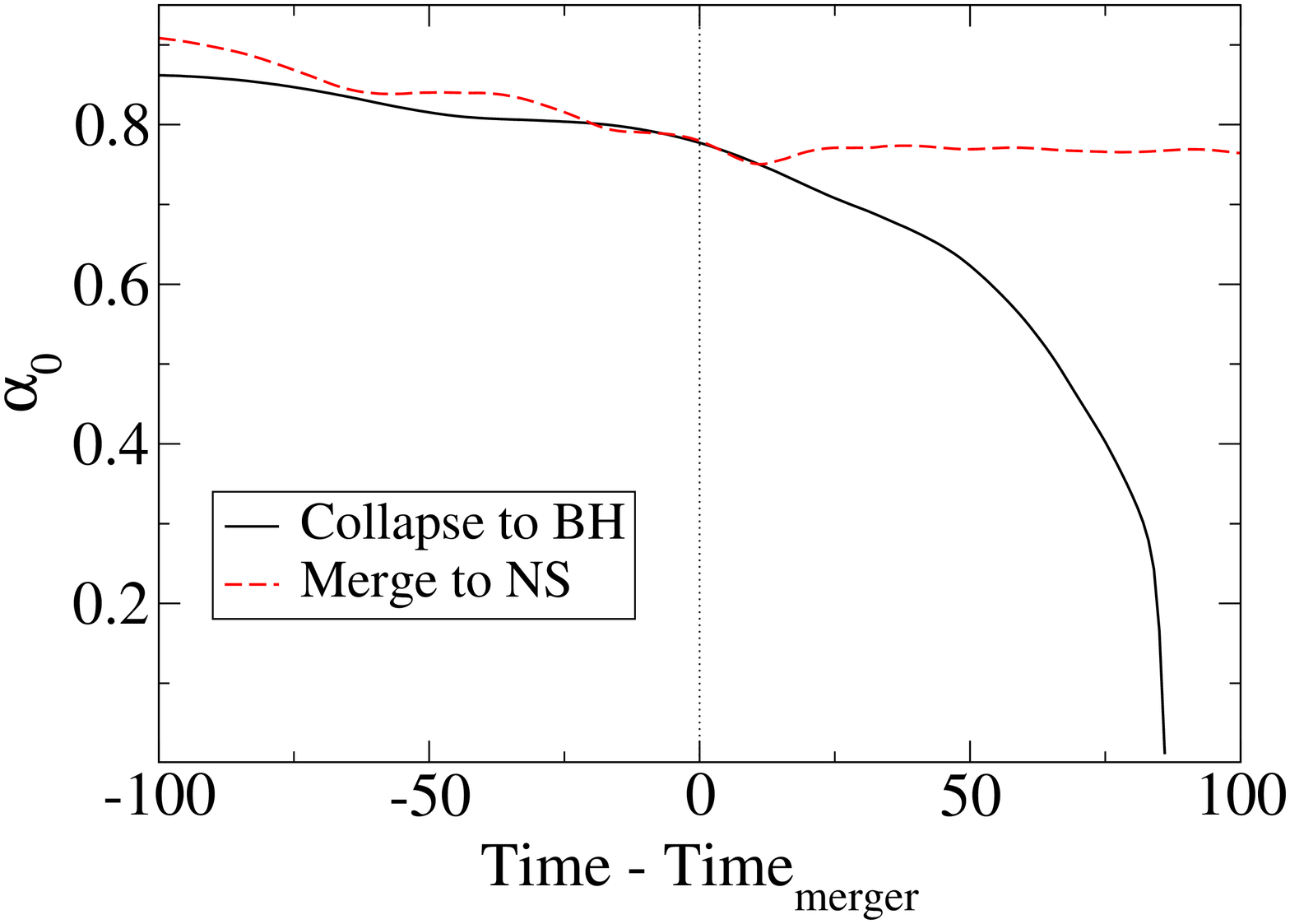,height=8.0cm,angle=270}
\caption{The lapse at the origin as a function of time for the orbiting
  polytropes and their merger to either a black hole or neutron star.  For
  the sake of comparison, we have defined $t_{\rm merger}$ to be
the instand at which the stars come into contact.
    See Figures~\ref{fig:lapsesnapshots},\ref{fig:NOBHlapsesnapshots} for 
    contour plots of the lapse in the two cases.
}
  \label{fig:minlapse}
\end{center} 
\end{figure}

Upon merger, the object's pressure and rotation can not support the star and it 
quickly collapses to a black hole. 
As described earlier, our simulations are carried 
out with harmonic slicing which is not singularity 
avoiding~\cite{Alcubierre:2002iq}. 
Although the lapse collapses to zero as illustrated in
Figs.~\ref{fig:minlapse} and \ref{fig:lapsesnapshots},
it does not collapse sufficiently fast to avoid numerical problems.
As a result, the size of the merged object decreases rapidly
and the code crashes when it can no longer resolve the
physical length-scales  within the allowed maximum refinement
levels as shown in the final frame of Figure~\ref{fig:fluidsnapshots}.
Ongoing work 
excises a region within a trapped surface to avoid 
this problem. We defer to future work a full analysis of the post-merger case
and the transition to a quasinormal ringing pattern in the 
radiation~\cite{bnstobh_us}.

%
\subsection{Differentially rotating neutron star}

\begin{figure}
\begin{center} 
\epsfig{file=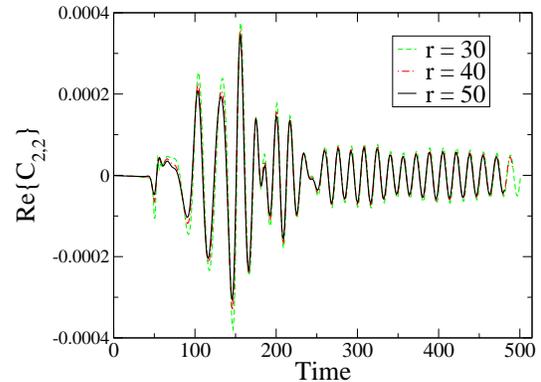,height=8.0cm,angle=270}
\caption{The merger waveform for the collision resulting in a single
  compact star extracted at three different stellar radii: 30, 40, and 50.
    The domain of the simulation is $\pm 152$ stellar radii.  After
the merger a transient behavior is observed. In particular, the features
at $t\simeq 180, 240$ result from marked oscillations in the produced
bar-like configuration (as seen in the co-rotating frame). Afterwards
the gravitational waves due to the spinning bar exhibit a clear frequency at
   $\simeq 12.8$~kHz.}
\label{fig:finalstar}
\end{center} 
\end{figure}

In the case where the individual stars are initially 
separated (in coordinate space) by $4R$
and boosted with a speed of 0.0825 the merger 
 does not give rise to a prompt
collapse to a black hole, rather it produces a single
differentially rotating star (See Figs.~\ref{fig:nobhfluidsnapshots}--\ref{fig:NOBHlapsesnapshots}).  As in the previous case, the 
initial orbital dynamics correspond to an eccentric inspiral trajectory.
But upon merger, the object's pressure and rotation are sufficient to support
a newly formed star.  The merged object has a bar-like structure that 
is spinning with a characteristic pattern frequency.  The real part of
the coefficient $C_{2,2}$ of $r \Psi_4$ for this evolution (shown
here in Figure~\ref{fig:finalstar}) carries a signature of the merger 
($t$ approximately
from 100 to 200) and of the resulting spinning bar ($t$ greater than 250).
Qualitatively, the outcome of this
evolution agrees with the results presented for the fully relativistic
simulation labeled ``E-1'' in~\cite{Shibata:2002jb}, and even with the results
from the post-Newtonian SPH simulation labeled ``F1'' 
in~\cite{faberrasio}; compare, 
for example, our Figure~\ref{fig:finalstar} 
with  Figure 11 in~\cite{Shibata:2002jb} 
and Figure 3 in~\cite{faberrasio}.  These two earlier 
simulations also followed the merger of equal-mass, initially irrotational
neutron stars having a $\Gamma=2$ equation of state.  However,
the bar-like structure survives noticeably longer in our 
simulation than in the evolution presented in~\cite{faberrasio}, 
and in our simulation
the radiation signature appears to carry more detail about the 
post-merger dynamics than in either of these earlier evolutions.
Specifically, the structure discernible in Figure~\ref{fig:finalstar}  
between the times
180 and 240 reflects the fact that, as viewed from the co-rotating
frame of the bar, the bar itself is experiencing nontrivial oscillations.

\begin{figure}
\begin{center} 
\epsfig{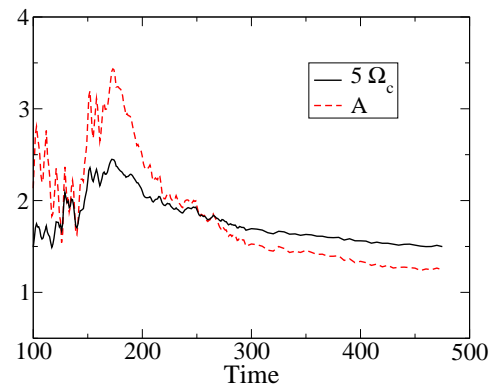}
\caption{The fitted values of $\Omega_c$ and $A$ as determined from
the fluid's tangential velocity. The merger takes place at about
$t\simeq 140$ after which the angular velocity rises during a transient
stage and then slowly decreases.}
\label{fig:fits}
\end{center} 
\end{figure}

The neutron star that forms from this merger is strongly differentially
rotating.  In an effort to quantify this, in the latter stages of
the evolution we fit the internal motions of the star to a rotation law 
of the form,
\begin{equation}
\Omega(r) = \frac{\Omega_c}{1+A r^2 \sin(\theta)^2}
\end{equation}
which has proven to be useful in numerous other investigations (see for instance, \cite{hachisu,Shibata:2002jb,Lyford:2002ip}). 
Figure~\ref{fig:fits}  shows the time-dependent behavior of
the fitted parameters $\Omega_c$ and $A$.  We note in particular
that the ratio of the central and near-surface value of $\Omega$
at the equator is $\Omega_c/\Omega_{\rm eq} \approx 0.34$.

%
%
\section{Conclusion}

Neutron stars will be important sources of gravitational waves for
the next generation of gravitational wave detectors. While waveforms
from neutron star binaries are weaker than those produced by
binary black holes due to the allowed neutron star masses, their signals
are expected to be richer, as the gravitational waves will also
carry information about the matter.  Indeed, gravitational waves are
expected to become an important probe of neutron star physics,
addressing questions such as the equation of state for nuclear
matter and the nature of progenitors for short, hard gamma-ray
bursts.

We have constructed a code that solves the combined Einstein and fluid equations in
three spatial dimensions, with no symmetry assumptions, and we use
\had\ for distributed AMR.  AMR is an essential element
of our method, as it allows us to place the outer boundaries far
from the binary, while the shadow hierarchy allows us to refine each
star individually without a priori assumptions about their motion.
We have carefully verified our numerical results by performing runs
at different resolutions, using grids with different physical outer
boundaries, extracting $\Psi_4$ at different radii, and varying the
floor applied to the fluid densities.  Moreover, we studied the
radial pulsation frequencies for a $\Gamma=2$ polytropic TOV star,
finding excellent agreement between our results and the expected
perturbative values.  The successful conclusion of these tests gives
us confidence in the physical results obtained from our code. 

As a first application of this code in a demanding scenario, we present
 a detailed study of two binary neutron
star mergers, one resulting in a final black hole and the other a
final neutron star.  In both cases we examine the gravitational
wave emission by extracting the $\ell=2$, $m=2$ mode of $r\Psi_4$.
$\Psi_4$ is extracted sufficiently far from the binary within the wave 
zone, and extraction is done at three different radii.  
In the first case, $\Psi_4$ is extracted up until the lapse collapses,
and in the second case the wave signal is extracted until a
final differentially rotating star is reached.
 A comparison
to a post-Newtonian analysis allows us to understand better the
gravitational wave signals and the orbital kinematics, such as
orbital trajectories, frequencies, and eccentricities.  For example,
the initial data describes an eccentric orbit.  The effect of
the eccentricity can be observed in the alternating pattern of
larger and smaller extrema in $\Psi_4$ as well as a modulation in
the observed wavelength. Both features are expected from a Post-Newtonian
analysis of an eccentric orbit.  The orbits circularize
through gravitational wave emission, and the solution around the 
time of collapse is largely spherically symmetric.  In the second case,
the neutron star merger results in a large strongly differentially rotating
star. The observed maximum density after
the merger does not lie at the origin but oscillates, in the co-rotating frame,
in a bar-like fashion in between $\simeq 0.2 R_{\rm final}$
and $\simeq 0.4 R_{\rm final}$ (with $R_{\rm final}$ the equatorial radius of the
merged object).

The work presented here raises additional questions that we will pursue
in a continuing research program.  
For example, we will continue to study the ringdown of the
final black hole formed in the first merger.  Studies of
the differentially rotating star formed in the second
case are continuing to determine whether this star eventually
collapses to form a black hole.
We will also examine a broader class
of initial data, including quasicircular and unequal mass binaries.
As mentioned 
previously, we also are investigating the effect of magnetic fields on 
the massive compact object formed in a merger and its possible subsequent 
collapse.  These results will be published in subsequent papers.

%
%

\renewcommand{\theequation}{A-\arabic{equation}}
\setcounter{equation}{0}
\section*{Appendix}\label{appendix}

It is customary in general relativity to adopt geometrized units $G=c=1$,
such that all quantities, including mass ($M$) and time ($T$),
have units of length ($L$).  Vacuum 
solutions are invariant under changes in this fundamental length scale $L$.
A quantity $X$ that scales as $L^lM^mT^t$ 
can be converted into geometrized units by multiplying with the
factor $c^t~(G/c^2)^m$.  After the conversion to geometrized units, 
$X$ scales as $L^{l+m+t}$.
Most equations of state
break this intrinsic scale-invariance, and the fundamental length-scale must
be fixed by additional choices. 
Once the new scale is chosen, transformations between 
geometrized and physical units can be easily made.
In the following, we summarize the basic procedure 
detailed in~\cite{Noblephd} to account for the proper scaling of
quantities. 

The polytropic EOS~(\ref{eq:polyEOS}) is specified by
the constants $\{ \kappa, \Gamma \}$, and the
quantities obtained  when using a particular set $\{ \kappa_1, \Gamma_1 \}$ 
can be scaled to those obtained using a second set 
$\{ {\kappa_2}, {\Gamma}_2 \}$ by the factor
\begin{eqnarray}\label{rescaling}
  \frac{L_1}{L_2} = \frac{ { {\kappa_1}}^{1/2({\Gamma_2}-1)} }
                   {{\kappa_2}^{1/2(\Gamma_2-1)}} \,  .
\end{eqnarray}
There are two common approaches in the literature to set this additional 
length scale.  The first one is obtained by fixing a constant physical 
quantity, e.g., the solar mass $M_\odot=1$, and from it deduce the appropriate 
conversion factors. That is, if a quantity
$\hat{X}$ has dimensions of $L^lM^mT^t$, its 
dimensionless counterpart, $X$, is obtained from  the following equation:
\begin{equation}\label{unit1}
   \hat{X} = \left(\frac{G~M_\odot}{c^2}\right)^{l+t} \frac{M^m_{\odot}}{c^t} 
    X\, .
\end{equation}
There is still the freedom to choose $\kappa$, and all dimensions
are scaled with this parameter. Usually the choice $\kappa=100$ is preferred
because it leads to physical units which are close to the current observations.
For instance, TOV stars constructed with these parameters have a maximum 
stable mass of $\hat{M}_{\rm max}=1.64 M_\odot$ with
a radius of $\hat{R}_{\rm max}=14.11$~km.

The second method for choosing the length scale is explained in detail 
in~\cite{Noblephd},
and is more involved.   It is based on fixing
the maximum stable mass for a family of solutions (with given 
$\{ \kappa=1, \Gamma \}$) to a physically motivated value. 
Thus, a quantity $\hat{X}$ with
dimensions  $L^lM^mT^t$ is obtained by using the relation:
\begin{equation}\label{unit2}
   \hat{X} = \hat{\kappa}^x c^y G^z X,
\end{equation}
where
\begin{eqnarray}
   x &=& \frac{l + m + t}{2 (\Gamma-1)} \, , \quad
   y = \frac{(\Gamma-2)l + (3\Gamma-4)m - t}{\Gamma-1}\, ,
\nonumber   \\
   z &=& -\frac{l + 3m + t}{2} \, .    
\end{eqnarray}
In this method $\hat{\kappa}$  has dimensions. We now identify
the maximum stable mass for the given polytrope to some physical maximum mass.
For a neutron star, the observed maximum mass is 
$\hat{M}_{\rm max} = 1.4 M_\odot$.
 
Although this second method for fixing the fundamental length scale generally
leads to different results from
the first, it can be checked that for $\Gamma=2$
both methods (the first one with $\kappa=100$, while the second one always has 
$\kappa=1$) provide the same scaling factors when the physical maximum stable 
mass is set
to $\hat{M}=1.64 M_\odot$. Since the dimensionless maximum stable
mass is $M=0.164$, Eq.~(\ref{unit2}) can be solved for $\hat{\kappa}$ with
$\{l=0,m=1,t=0\}$, giving
$\hat{\kappa}=1.456 \times 10^5 {\rm cm}^5/\left({\rm g}\, {\rm s}^2\right)$. With this value,
(\ref{unit2}) can again be used to recover the dimensions of any quantity.

%
%
\begin{acknowledgments}
We would like to thank J. Frank, J. Pullin, I. Olabarrieta and O. Reula for
stimulating discussions. 
This work was supported by the National Science Foundation under grants
PHY-0326311, PHY-0554793, AST-0407070 and AST-0708551 to Louisiana State 
University, PHY-0326378 and PHY-0502218
to Brigham Young University, and PHY-0325224 to Long Island University.
This research was also supported in part by the National Science Foundation
through TeraGrid resources provided by SDSC under allocation award PHY-040027.
In addition to TeraGrid resources, we have employed clusters belonging
to the Louisiana Optical Network Initiative (LONI), and clusters
at LSU (mike) and BYU (marylou4).
\end{acknowledgments}

%
%
\bibliography{./tov}
\bibliographystyle{apsrev}

%
%
\end{document}